\documentclass[11pt]{article}

\usepackage{amsmath,amsfonts,amsthm}
\usepackage[breaklinks,pdfpagemode=None,pdfview=FitH,pdfstartview=FitH,citebordercolor={0 0 1},linkbordercolor={0 0 1},urlbordercolor={0 0 1},pagebordercolor={0 0 1}]{hyperref}

\input colordvi    

\newtheorem{theorem}{Theorem}[section]
\newtheorem{proposition}[theorem]{Proposition}
\newtheorem{corollary}[theorem]{Corollary}
\newtheorem{lemma}[theorem]{Lemma}

\renewcommand{\a}{\alpha} 
\newcommand{\e}{\varepsilon}
\newcommand{\f}{\varphi}
\newcommand{\D}{\Delta}
\newcommand{\G}{\Gamma}
\renewcommand{\O}{\Omega}

\renewcommand{\AA}{\mathcal{A}}

\newcommand{\GG}{\mathcal{G}}

\newcommand{\PP}{\mathcal{P}}

\newcommand{\Scal}{\mathcal{S}}
\newcommand{\ZZ}{\mathcal{Z}}

\newcommand{\bR}{\mathbf{R}}
\newcommand{\bZ}{\mathbf{Z}}

\begin{document}

\title{Finite range Decomposition of Gaussian Processes}

\author{
    David C. Brydges\thanks{Supported by NSERC}\\
    The University of British Columbia\\
    1984 Mathematics Road\\
    Vancouver, B.C., Canada V6T 1Z2
\and
    G. Guadagni\\
    Department of Mathematics\\
    University of Virginia\\
    Charlottesville VA 22904-4137
\and
     P. K. Mitter\thanks{Laboratoire Associ\'e au CNRS, UMR 5825}\\
     Laboratoire de Physique Math\'ematique\\ 
     Universit\'e Montpellier 2\\
     Place E. Bataillon, Case 070\\ 
     34095 Montpellier Cedex 05 France
}    

\date{ {\it In Honour of G. Jona-Lasinio}  \\
\today
}

\maketitle

\begin{abstract}
Let $\D$ be the finite difference Laplacian associated to the
lattice $\bZ^{d}$.  For dimension $d\ge 3$, $a\ge 0$ and $L$ a
sufficiently large positive dyadic integer, we prove that
the integral kernel of the resolvent $G^{a}:=(a-\D)^{-1}$ can be
decomposed as an infinite sum of positive semi-definite functions $
V_{n} $ of finite range, $ V_{n} (x-y) = 0$ for $|x-y|\ge
O(L)^{n}$. Equivalently, the Gaussian process on the
lattice with covariance $G^{a}$ admits a decomposition into
independent Gaussian processes with finite range covariances. For
$a=0$, $ V_{n} $ has a limiting scaling form $L^{-n(d-2)}\Gamma_{
c,\ast }{\bigl (\frac{x-y}{ L^{n}}\bigr )}$ as $n\rightarrow
\infty$. As a corollary, such decompositions also exist for fractional
powers $(-\D)^{-\alpha/2}$, $0<\alpha \leq 2$.

The results of this paper give an alternative to the block spin
renormalization group on the lattice.
\end{abstract}

\vskip0.3cm
{\bf keywords:}\ gaussian processes, finite range decomposition, lattice,
renormalization group, L\'evy processes.

\section{Introduction} \label{sec-introduction}
\setcounter{equation}{0}

A smooth Gaussian process $\zeta (x)$ on ${\bf R}^{d}$
with the property that the expectation $ E\zeta (x)\zeta (y) = 0 $ when
$|x-y|\ge L$ will be said to have {\it finite range} $L$.  What is the
class of Gaussian processes $\phi$ that can be expressed as a sum
$\phi = \sum_{j}\zeta_{j}$ of independent finite range processes with
ranges $ \sim L^j$ for some $L$? Let us call such processes {\it
finite range decomposable}.

We can reformulate this in terms of the covariance: a Gaussian process
$\phi$ is finite range decomposable if the covariance $C (x,y):=E\phi
(x)\phi (y)$ can be written as a sum $C = \sum_{j}V_{j}$ where each
$V_{j} (x,y)$ is positive semi-definite and has finite range $\sim
L^{j}$.  In this form the question has already received a partial
answer in the study of ground states for many-body Hamiltonians.  In
particular, in \cite{HaiSei2002} Hainzl and Seiringer discuss this
background and consider the decomposition
\begin{equation}\label{eq-intro-1}
V (x) = \int_{0}^{\infty} dr \ g (r) \chi_{r/2}\ast \chi_{r/2} (x)
\end{equation}
of a radial function $V (x)$ as a weighted integral of tent functions
$\chi_{r/2}\ast \chi_{r/2} (x)$, where $\chi_{r/2}$ is the indicator
function of the ball of radius $r/2$.  An explicit formula for $g$ in
terms of $V$ is derived.  For example, in three dimensions,
\[
g (r) = -\frac{2}{\pi } ( V'' (r)/r)'
\]
so necessary and sufficient conditions for $g\ge 0$ in terms of $V$
are readily formulated. In particular Coulomb and Yukawa potentials in
three dimensions have decompositions with nonnegative $g$.

This is relevant to our question because the tent function is positive
semi-definite and therefore, when $g (r) \ge 0$ and $I$ is an interval
$[a,b)$,
\[
V_{I} (x) := \int_{I} dr \ g (r) \chi_{r/2}\ast \chi_{r/2}
(x)
\] 
is also positive semi-definite. By breaking up the range of the $r$
integration in (\ref{eq-intro-1}) into a disjoint union of intervals,
$I_{j}:=[L^{j},L^{j+1}), j \in \bZ$, we have $V = \sum V_{I_{j}} $
and there is a corresponding finite range decomposition $\phi =
\sum_{j}\zeta_{j}$ when $\phi$ is the Gaussian process with covariance
$V (x-y)$ with $g (r) \ge 0$ and $\zeta_{j}$ has covariance
$V_{I_{j}}$.

These decompositions are not the final answer to our question, because
we are also interested in kernels defined on the lattice $\bZ^{d}$ and
furthermore one may get a wider class by not insisting on
decompositions based on tent functions.  For lattices or the continuum
we have preliminary results that suggest that resolvents of quite
general elliptic operators and fractional inverse powers of elliptic
operators are candidates for such decompositions.

Our interest in this question is rooted in the Renormalization Group
(RG). In quantum field theory and other contexts the RG is a method to
calculate the expectation $E\ZZ$ of a functional $\ZZ = \ZZ (\phi)$ of
a Gaussian field $\phi $. One decomposes $\phi =\sum_{j\ge
1}\zeta_{j}$ as a sum of independent Gaussian fields $\zeta_{j}$ and
integrates out each $\zeta_{j}$ one at a time. Let $E_{j}$ be the
expectation that integrates out $\zeta_{j}$.  Then the RG is the
sequence of maps $\ZZ_{j} \mapsto \ZZ_{j+1}:=E_{j+1}\ZZ_{j}$.  $E\ZZ$
is obtained from $E\ZZ = \lim \ZZ_{j}$, starting with $\ZZ_{0}:=\ZZ$.
[These ideas are explained further in
Section~\ref{sec-probability}]. The point is to choose the
decomposition to have special properties so that each expectation
$E_{j}$ is more amenable to analysis than the whole expectation $E$
and furthermore so that the map $\ZZ_{j} \mapsto \ZZ_{j+1}$ can be
analysed within the context of dynamical systems.  In particular the
RG is very informative when the limiting map is autonomous, up to a
scaling. This is possible when the covariances $V_{j}$ of $\zeta_{j}$
are becoming \emph{self-similar}. This means that there should exist a
{\it dimension} $[\phi]$ such that $V_{j} (x,y)=:
L^{-2j[\phi]}\Gamma_{j}(L^{-j}x,L^{-j}y)$ defines a scaled covariance
$\Gamma_{j}$ which tends to a limit as $j\rightarrow \infty$.

When the covariances $V_{j}$ are finite range, the map $\ZZ_{j}
\mapsto \ZZ_{j+1}$ can be studied by using the independence of
$\zeta_{j} (x)$ and $\zeta_{j} (y)$ for $|x-y|\geq L^{j}$.  We amplify
on this remark at the end of this introduction. In some ways, the use
of these finite range covariances gives the simplest framework that
goes beyond the hierarchical models.  This program is also close to
the technique that was invented by Fr\"ohlich and Spencer in
\cite{FrSp81} to study the Kosterlitz-Thouless transition, but it is
in principle more precise and more robust.

Let $\Delta$ be the finite difference Laplacian associated to the
lattice $\bZ^{d}$.  We will consider decompositions for the kernel
$G^{a} (x-y)$ of the resolvent $(a-\Delta)^{-1}$ and we will also
consider the Green's function of a stable L\'evy process
which is the kernel $C (x-y)$ of $(-\Delta)^{-\alpha /2}$
where $0<\alpha < 2$. The mass parameter $a \in
[0,\infty )$. We state our results for dimensions $d \ge 3$ because
lower dimensions require extra discussions for the case $a=0$, but the
basic construction is valid in lower dimensions as well.  $L$ is a
parameter of the form $2^{p}$. $p$ can be any sufficiently large
integer.  There are many technical results in this paper so we have
summarised the main points in the following theorem which is a
combination of results from the theorems in the rest of the paper.

\begin{theorem}\label{thm-intro-1} For $n=0,1,2\dotsc $, and all $a\ge 0$,
there are
positive semi-definite functions $\Gamma_{n}^{a} (x)$ defined for $x
\in (L^{-n}\bZ)^{d}$ such that 
\begin{enumerate}
\item $G^{a} (x-y) = \sum L^{-2n [\phi]} \
\Gamma^{L^{2n}a}_{n}
\bigg(
\frac{x-y}{L^{n}}
\bigg)$ with $[\phi]= (d-2)/2$
\item $\Gamma_{n}^{a} (x)=0$ for $|x|\ge 6L$
\item $|\hat{\Gamma}_{n}^{a} (p) |\le c_{k,L} (1+|p|)^{-k}$ for $p \in
[-L^{n}\pi ,L^{n}\pi ]^{d}$
\item $\hat{\Gamma}_{c,\ast}^{a} (p):=\lim_{n\rightarrow
\infty}\hat{\Gamma}_{n}^{a} (p) $ exists pointwise in $p$
\item Fix a positive integer $l$ and let $\epsilon
=L^{-l}$.  Then  $\Gamma_{c,\ast}^{a} = \lim_{n\rightarrow\infty
}\Gamma_{n}^{a}$ exists in $L^{\infty } ((\epsilon
\bZ)^{d})$. Furthermore the $L^{\infty}$ limit of any multiple lattice
derivative of $\Gamma_{n}^{a} $ also exists and is the corresponding 
continuum derivative of $\Gamma_{c,\ast}^{a} $
\item Analogous statements hold for $C (x-y)$, but with $[\phi]=
(d-\alpha)/2 $.
\end{enumerate}

\end{theorem}

The multiscale expansion in (1) above is obtained in (\ref{eq-decomp.31})
of section 3, and the finite range property (2) is given in 
Lemmas \ref{lem-decomp.1}, \ref{lem-decomp.2}. 
The bound in (3) follows from Theorem \ref{thm-bounds.5} of section 5. 
(4) above is obtained in section 6 in the course of proving Theorem 
\ref{thm-convergence.1}, see (\ref{eq-converge.124}). (5) above is part of
Theorem \ref{thm-convergence.1}. Turning to (6) above, the finite range
decomposition of $C$ is obtained in section 4 ( see (\ref{eq-prob.32}),
(\ref{eq-prob.325}) et seq.). A uniform bound on L\'evy fluctuation
covariances is supplied in section 5 ( see Corollary \ref{cor-bounds.6} ).
The statement analogous to (5) above is part of 
Corollary \ref{cor-convergence.2} of section 6.
\vskip0.3cm

Let us call the Gaussian fields in the decomposition \emph{fluctuation
fields}. Other (wavelet) decompositions were developed in the context
of the {\it Block Spin } Renormalization Group of Kadanoff and Wilson
by Gawedzki and Kupiainen, \cite{GaKu80,GaKu83} as well as, in related
work, by Balaban, \cite{Bal82a,Bal82b}.  Although the
Gawedzki-Kupiainen fluctuation fields are not finite range, they have
their own advantages: notably they are independent lattice fields,
determined by random variables defined on increasingly coarse lattices
of spacing $L^{n}$.  Our decomposition achieves this only in the
weaker sense that, with high probability, $\zeta_{n} (x)\approx
\zeta_{n} (y)$ for $|x-y|\ll L^{n}$, but $\zeta_{n}$ retains low
probability variations on all smaller scales.  This is a price paid
for retaining translation invariance on small scales.

In section 3 the aforementioned finite range decompositions are
obtained. The rescaled fluctuation covariances live on finer and finer
lattices, but all have the same finite range.  In section 4
probabilistic aspects of our construction are discussed and the finite
range decomposition of the L\'evy Greens function is obtained.  We are
interested in the L\'evy Greens function $(-\Delta)^{-\alpha /2}$
because varying the parameter $\alpha$ affects the scaling of the
associated field and gives insight into the dynamical system
$\ZZ_{j}\mapsto \ZZ_{j+1}$. Moreover it is of intrinsic interest in 
various problems in probability theory. For example we may wish to study
critical properties of self avoiding L\'evy walks by renormalization group
methods. We also discuss in this section renormalization group 
transformations based on the above finite range decompositions.  

Section 5 is devoted
to bounds. Here our main result is Theorem~\ref{thm-bounds.5}. This
theorem states that every member of the sequence of rescaled
fluctuation covariances is uniformly bounded in lattice Sobolev norms
of arbitrarily high degree. The bound is independent of the lattice
spacing. Finally in Section 6 we prove that the sequence converges in
Sobolev norms in a precise sense to its continuum limit which is
appropriately identified. The continuum limit is smooth. This is the
content of our main Theorem~\ref{thm-convergence.1}.

We conclude with a brief indication of the role of the finite range property
in the analysis of $\ZZ_{j}\mapsto \ZZ_{j+1}$.  Consider
$\ZZ_{0} \mapsto \ZZ_{1}$ when 
$$
\ZZ_{0} = \ZZ_{0} (\Lambda,\phi_{0}):=
\prod_{x \in \Lambda }e^{-\lambda \phi_{0}^{4} (x)}
$$
where $\Lambda \subset \bZ^{d}$ is a large box shaped subset of
lattice points which is a disjoint union of some standard cube shaped
subsets of lattice points $\Delta \subset \bZ^{d}$ of side greater
than the range of the first field $\zeta_{1}$ in a finite range
decomposition $\phi_{0} = \sum_{j\ge 1}\zeta_{j}$. Let $\phi_{1}:=
\sum_{j\ge 2}\zeta_{j}$, $X \subset \bZ^{d}$ and
\[
        \tilde{\ZZ}_{1} (X)= \prod_{x\in X}e^{-\lambda \phi_{1}^{4} (x)}
\]
Note that $\tilde{\ZZ}_{1} (X)$ is independent of $\zeta_{1}$.  Then
$$
\ZZ_{1} (\Lambda) = E_{1}\prod_{\Delta \subset \Lambda}
\ZZ_{0} (\Delta) = E_{1}\prod_{\Delta \subset \Lambda}
\bigg(
\ZZ_{0} (\Delta)-\tilde{\ZZ}_{1} (\Delta)+\tilde{\ZZ}_{1} (\Delta)
\bigg)
$$
Write $\delta \ZZ_{0}(\Delta) := \ZZ_{0} (\Delta)-\tilde{\ZZ}_{1}
(\Delta)$.  The product expands into a sum over $X$ of terms
\[
\prod_{\Delta \not \subset X}\tilde{\ZZ}_{1} (\Delta) 
\prod_{\Delta \subset X}\delta \ZZ_{0}(\Delta) 
\]
In other words, $X$ labels the factors where $\delta \tilde{\ZZ}_{0} $ is
selected. We can partition $X$ into disjoint connected components
$X_{1},\dotsc X_{M}$, where $X_{i}$ is connected when the cubes
$\Delta$ in $X_{i}$ are such that one can pass between any pair of
cubes by a path whose steps are nearest neighbour cubes in $X_{i}$.
Let 
$$K_{0} (X_{i}):= \prod_{\Delta \subset X_{i}} \delta
\ZZ_{0}(\Delta)
$$
Then
$$
\ZZ_{1}(\Lambda) = \sum \frac{1}{ M!}\sum_{X_{1},\dots ,X_{M}}
\tilde{\ZZ}_{1} (X_{0})
E_{1}\prod_{i}K_{0} (X_{i})
$$
where the connected sets $X_{j}$ are disjoint and $X_{0}:=\Lambda
\setminus \cup_{i\ge 1} X_{i}$. Notice that $\tilde{\ZZ}_{1}$ can be
and has been moved outside the expectation since it is independent of
$\zeta_{1}$. Now comes the key point: The sets $X_{j}$ are 
connected unions of nearest neighbour cubes of side length greater
than the range of $\zeta_{1}$. Since they are disjoint they are separated 
by a distance greater than the range of $\zeta_{1}$. Therefore, 
by the finite range property, 
\begin{equation}\label{eq-intro-2}
\ZZ_{1}(\Lambda) = \sum \frac{1}{ M!}\sum_{X_{1},\dots ,X_{M}}
\tilde{\ZZ}_{1} (X_{0})
\prod_{i}E_{1}K_{0} (X_{i})
\end{equation}
The perturbation on a large volume $\Lambda$ has been reduced to local
calculations $ E_{1}K_{0} (X_{i}) $ and the standard but heavy machinery of
cluster expansions is being replaced by independence and geometry.

Unlike $\ZZ_{0} (\Lambda)$ the image functional $\ZZ_{1} (\Lambda)$ no
longer factors into contributions from boxes, which is the great
simplification of hierarchical models, but there is still a large part
$X_{0}$ of $\Lambda$ where this property is retained. For the next RG
map one proves that the more general form (\ref{eq-intro-2}) is stable
in the sense that $\ZZ_{2}$ can also be written in the same form but
with a different $K_{1}$ and with larger cubes $L\Delta $.  The
program is then to prove that $E_{j}K_{j} (X)$ gives very little
weight to connected sets $X$ which are unions of many boxes
$L^{j}\Delta$.  This can be facilitated by making a better choice of
$\tilde{\ZZ}_{1}$, since the derivation is valid for other choices of
$\tilde{\ZZ}_{1}$. In particular, one can replace $\lambda$ in
$\tilde{\ZZ}_{1}$ by some other value $\lambda_{1}$ chosen to minimise
$E_{1}K_{0}$.  This idea leads to a flow of the coupling constant
$\lambda \rightarrow \lambda_{1}$.
 
The first use of finite range covariances was in \cite{MiSc2000}.  
\cite{BMS03} is another appearance of finite range covariances.

\section{Preliminaries}\label{sec-preliminaries}
 \setcounter{equation}{0}

Throughout we will assume that $d\ge 3$. Let $L$ be a large integer power of
$2$. Define $\e_{n} = L^{-n}$.
We will be working on a sequence
of lattices $(\e_{n}\bZ)^{d}\subset {\bf R}^{d}$, 
$(\e_{n}\bZ)^{d}\subset (\e_{n+1}\bZ)^{d}$, with $n=0,1,2,...$ and 
eventually passing
to ${\bf R}^{d}$.
$(\e_{n}\bZ)^{d}$  is equipped with the discrete topology. The
measurable sets are subsets of points and the measure $dz$ on $(\e_{n}\bZ)^{d}$ 
is defined by
\begin{equation}
\int_{(\e_{n}\bZ)^d}dz\ \ f(z)= \e_{n}^{d}\sum_{z\in
(\e_{n}\bZ)^d}f(z)\label{eq-prel.1}
\end{equation}
We endow ${\bf R}^{d}$ with the distance function
\begin{equation}
\vert x-y\vert = {\rm max}_{1\le j\le d}\ \vert x_{j}-y_{j}\vert
\label{eq-prel.21}
\end{equation}
Let 
\begin{equation}
U(R)=\bigl (-\frac{R}{ 2},\frac{R}{ 2}\bigr )^{d}\subset {\bf R}^{d}
\label{eq-prel.22}
\end{equation}
be an open cube of edge length $R$. Define
\begin{equation}
U_{\e_{n}}(R) = U(R)\cap (\e_{n}\bZ)^{d} \label{eq-prel.2}
\end{equation}
and its boundary
\begin{equation}
\partial U_{\e_{n}} = \{y \not \in \> U_{\e_{n}} :\> |x-y|=\e_{n},\>
{\rm some} \> x\in U_{\e_{n}}\} \label{eq-prel.3}
\end{equation}
The distance function $ |.|$ is that induced from ${\bf R}^{d}$. We
denote by ${\bar U_{\e_{n}}}= U_{\e_{n}}\cup\partial U_{\e_{n}} $ the
closure of $U_{\e_{n}} $.
The lattice Laplacian $\Delta_{\e_{n}}$ is defined by the
quadratic form
\begin{equation}
(f,-\Delta_{\e_{n}}f)_{L^{2}((\e_{n}\bZ)^d)}=\e_{n}^{d}\sum_{<x,y>}
\e_{n}^{-2}|f(x)-f(y)|^{2} \label{eq-prel.4.5}
\end{equation}
where the sum runs over the nearest neighbour points in $(\e_{n}\bZ)^d$.
Let $\Scal = \{\hat{e}_{1},...,\hat{e}_{d} \}$ be the set of standard unit basis
vectors in $\bZ^{d}$. For any such lattice unit 
vector $ {\underline e} \in \Scal$ define forward  and backward
lattice derivatives $\nabla_{\pm \underline e}$ of a
function  in the direction $\underline e$ by
\begin{equation}
(\nabla_{\pm \underline e}f)(x)= \e_{n}^{-1}(f(x\pm{\e_{n}}{\underline e})-f(x)).
\label{eq-prel.4a}
\end{equation}
The backward derivative is defined so as to be the adjoint of the forward derivative.
Then the definition (\ref{eq-prel.4.5}) for the lattice Laplacian can
be written as
\begin{equation}
(f,-\Delta_{\e_{n}}f)_{L^{2}((\e_{n}\bZ)^d)}=\e_{n}^{d}\sum_{x,\underline
e\in \Scal} |(\nabla_{\underline e}f)(x)|^2 \label{eq-prel.4b}
\end{equation}
The corresponding resolvent  $G^{a}_{\e_{n}}$ with
\ \ $a\ge 0$ is
\begin{equation}
G^{a}_{\e_{n}}(x-y)=(-\Delta_{\e_{n}} + a)^{-1}(x-y) \label{eq-prel.5}
\end{equation}
\begin{equation}
=\int_{[-\pi/\e_{n},\pi/\e_{n}]^{d}}\frac{d^{d}p}{ (2\pi)^{d}}\>
\frac{e^{ip.(x-y)} }{a-\hat{\D}_{\e_{n}} (p)} \label{eq-prel.6}
\end{equation}
where
\begin{equation}\label{eq-prel.8}
\hat{\D}_{\e_{n}}(p) = 2\e_{n}^{-2}\sum_{\mu =1}^{d}\big(
\cos(\e_{n} p_{\mu})-1\big)
\end{equation}

\section{Multiscale Decomposition of the Resolvent}\label{sec-decomp}
\setcounter{equation}{0}

We say that a function $f (x,y)$ has \emph{finite range} $R$ if
\[
f (x,y)=0 \text{ for } |x-y|\ge R.
\] 
Consider the resolvent in $\bZ^d$ 
\begin{equation}
G^{a}(x-y)=(-\Delta + a)^{-1}(x-y) \label{eq-prel.7}
\end{equation} 
with $a\ge 0$.  We will first develop a multiscale decomposition for
the resolvent $G^{a},\ \ a\ge 0$ into smooth finite range positive
semi-definite functions.

As in (\ref{eq-prel.22}) and (\ref{eq-prel.2}), 
$U(L)\subset {\bf R}^{d}$ is an
open cube of edge length $L$ in ${\bf R}^{d}$
and $U_{\e}(L)=U(L)\cap (\e
\bZ)^{d}$ the induced cube in $(\e \bZ)^d$ with $\partial U_{\e}(L)$ its
boundary. Assume that the cube is centered at the origin. We will
suppress the argument $L$ when there is no risk of confusion.

On the lattice there is no need to distinguish functions from
measures. Nevertheless we use measures in cases where the associated
continuum object is a measure. A case in point is the
\emph{lattice Poisson kernel} $\PP^{a}_{U_{\e}}(x,du)$, which by definition
is the measure supported on $\partial U_{\e}$ such that
\begin{equation}\label{eq-decomp.660}
h (x)=\PP^{a}_{U_{\e}}(x,f):=\int {\PP}^{a}_{U_{\e}}(x,du)f(u)
\end{equation}
is the unique solution to the boundary value problem
\begin{equation}
(-\Delta_{\e}+ a)h(x)=0 : \ \ x\in U_{\e} \label{eq-decomp.661}
\end{equation}
\begin{equation}
h(x)= f(x) :\ \ x\in \partial U_{\e} \label{eq-decomp.662}
\end{equation}  
where $f : \partial U_{\e}\rightarrow \bf R $. Existence and
uniqueness are easily proved since $h$ solves a finite dimensional set
of linear equations.  Note that because $a\ge 0$ a solution $h(x)$
satisfies the weak maximum principle. In Section~\ref{sec-probability}
we will see an explicit construction of which shows that
$\PP^{a}_{U_{\e}}$ is a \emph{defective} probability
measure. Defective means that the mass is at most one. (For $a=0$ and
$f=1$, $h\equiv 1$ which implies $\PP^{a}_{U_{\e}}(x,1)$ is a
probability measure). We will say $h$ is \emph{$a$-harmonic} in
$X_{\e}$ if $h$ solves $(-\Delta_{\e}+ a) h (x)=0$ 
in $X\cap (\e \bZ)^{d}$.

Let g(x) be a rotationally invariant non-negative $C^{\infty}({\bf R}^{d})$ 
function of compact support such that 
\begin{equation}
g(x)=0 :\> |x|\ge \frac{L}{ 4} \label{eq-decomp.7a}
\end{equation}
with the normalization
\begin{equation}
\int_{{\bf R}^{d}} dx \> g(x)=1 \label{eq-decomp.7b}
\end{equation}
We restrict $g(x)$ to the lattice $(\e \bZ)^d$ and choose the normalization 
constant $c_{\e}$ such that 
\begin{equation}
\int_{(\e \bZ)^d} dx \> c_{\e}g(x)=1 \label{eq-decomp.8}
\end{equation}
Since $g(x)$ is a continuous function of compact support Riemann sums
converge. Hence $c_{\e}$ is a continuous function of $\e$ on the
compact set $0\le \e \le 1$ and thus uniformly bounded. Moreover
$c_{\e}\rightarrow 1$ as $\e \rightarrow 0$.  

Now comes the main idea.  The point of the function $g$ is to
avoid needing detailed knowledge of the Poisson kernel for the
lattice.  We are about to use the Poisson kernel to define an
averaging operator that leaves a-harmonic functions unchanged.  This
property leads to the finite range property in
Lemma~\ref{lem-decomp.1}.  It is relatively easy to prove that our
averaging operator is smoothing (uniformly in lattice spacing) because
when checking differentiability, derivatives either fall on $g$ which
is smooth by choice or on the $x$ argument of $\PP^{a}_{U_{\e}}(x,du)$
with $x$ forced to be away from the boundary $\partial U_{\e}$ so that
the easy part of standard elliptic techniques is sufficient to prove
smoothness uniformly in the lattice spacing.

Given a function $f:(\e \bZ)^{d} \rightarrow \bf R$ we define the {\it
averaging map} :
$$f\rightarrow A^{a}_{\e}(L)f$$
where
\begin{equation}
(A^{a}_{\e}(L)f)(x)=\int_{(\e \bZ)^d} dz \> c_{\e}g(z-x)\int
\PP^{a}_{U_{\e}(L)}(x-z,du)f(u+z) \label{eq-decomp.9}
\end{equation}
Note that this can also be written as
\begin{equation}
(A^{a}_{\e}(L)f)(x)=\int_{(\e \bZ)^d} dz \>
c_{\e}g(z-x)\PP^{a}_{U_{\e}(L,z)}(x,f) \label{eq-decomp.10}
\end{equation}
where $U_{\e}(L,z)$ is the translate of the cube $U_{\e}(L)$ so that
its center is now $z$ .  In summing over the translates we
have put in the smooth function $g$ , and not the delta function,
because we will need to take derivatives with respect to $x$ and this
is hard to do if $x$ is the center of the cube.  

Now this integration over all translates makes $A^{a}_{\e}(L)$
\emph{translation invariant}.  Translation invariance plays an
essential role in the proof of positive semi-definiteness of the
fluctuation covariance constructed below, (see
Lemma~\ref{lem-decomp.1}).  Proof: For $b\in (\e \bZ)^{d}$, let $f_{b}
(x) = f (x-b)$.  By change of variables $z\rightarrow z- b$ in
(\ref{eq-decomp.9})
\begin{equation}\label{eq-decomp.102}
(A^{a}_{\e}(L)f)(x-b) = (A^{a}_{\e}(L)f_{b})(x).
\end{equation}
>From (\ref{eq-decomp.9}) we see that for every fixed $x$,
$(A^{a}_{\e}(L)f)(x)$ defines a bounded, positive linear functional on
$C_{0}((\e \bZ)^{d})$, the space of functions of compact support on $(\e
\bZ)^{d}$. We have
$$|(A^{a}_{\e}(L)f)(x)|\le \Vert f \Vert_{\infty}$$
so that the norm of this linear functional is $\le 1$. This gives a family 
of defective probability measures $A^{a}_{\e}(L)(x,du)$ on $(\e \bZ)^{d}$ :
\begin{equation}\label{eq-decomp.1021}
(A^{a}_{\e}(L)f)(x)=\int_{(\e \bZ)^{d}}A^{a}_{\e}(L)(x,du) f(u)\end{equation}
The Fourier transform of this measure 
\[
\hat{A}^{a}_{\e} (p) = \int_{(\e \bZ)^{d}}\, du\, A^{a}_{\e}(L)
(0,du) e^{-ip.u}
\]
satisfies
\begin{equation}\label{eq-decomp.105}
|\hat{A}^{a}_{\e} (p)| \le \int_{(\e \bZ)^{d}}\, du\, A^{a}_{\e}(L)
(0,du) \le 1
\end{equation}

Define the {\it fluctuation covariance}
\begin{equation}
\Gamma^{a}_{\e}(x-y) =\> G^{a}_{\e}(x-y)-(A^{a}_{\e}(L)G^{a}_{\e}
A^{a}_{\e}(L)^{*} )(x-y) \label{eq-decomp.12}
\end{equation}
where by definition
\begin{equation}
(A^{a}_{\e}(L)G^{a}_{\e}A^{a}_{\e}(L)^{*} )(x-y) =\int\int
A^{a}_{\e}(L)(x,du)G^{a}(u-v)A^{a}_{\e}(L)(y,dv) \label{eq-decomp.13}
\end{equation}
The latter is the analogue of the {\it block spin covariance}
in statistical mechanics, \cite{GaKu86}. 
\begin{lemma}\label{lem-decomp.1}
$\Gamma^{a}_{\e}$ and $A^{a}_{\e}(L)G^{a}_{\e}{A^{a}_{\e}(L)}^{*}$
are positive semi-definite. $\Gamma^{a}_{\e}$ has finite range,
\begin{equation}
\Gamma^{a}_{\e}(x-y)=0\>:\> |x-y| \ge 3L \label{eq-decomp.14}
\end{equation}
and the Fourier transform $\hat{\Gamma}^{a}_{\e} (p)$ is
continuous in $p$ including at $p=0$, uniformly in $\e$.
\end{lemma}

\begin{proof} First we prove the finite range property. By the definition of
the Poisson kernel, if $f$ is $a$-harmonic in $x+U_{\e}$, then
${\PP}^{a}_{U_{\e}}(x,f)=f (x)$. Since $c_{\e}g$ in the definition of
$A^{a}_{\e}(L)$ was chosen to be a probability density with support in
$U_{\e} (L/4)$, for $f$ $a$-harmonic in $x+U_{\e} (5L/4)$,
\[
(A^{a}_{\e}(L)f)(x)=\int_{(\e \bZ)^d} dz \>
c_{\e}g(z-x)\PP^{a}_{U_{\e}(L,z)}(x,f) 
= \int_{(\e \bZ)^d} dz \>
c_{\e}g(z-x)f (x) = f (x) 
\]
When $|x-y| \ge 3L$, $x+U_{\e} (5L/4)$ and $y+U_{\e} (5L/4)$ are
disjoint.  Therefore $G^{a}_{\e}(u-v)$ is $a$-harmonic in each
argument in the appropriate region and therefore
$$(A^{a}_{\e}(L)G^{a}_{\e}A^{a}_{\e}(L)^{*} )(x-y) =G^{a}_{\e}(x-y)$$
which proves (\ref{eq-decomp.14}). Now we prove positive definiteness.
By translation invariance, we can take the fourier transform of
$\Gamma^{a}_{\e}(x-y)$ to get
\begin{equation}\label{eq-decomp.165}
\hat{\Gamma}^{a}_{\e}(p)=(1-|{\hat A^{a}_{\e}}(p)|^{2}){\hat
G^{a}_{\e}}(p)
\end{equation}
Now ${\hat G^{a}_{\e}}(p)\ge 0$ and $|{\hat A^{a}_{\e}}(p)|\le 1 $
where we have used (\ref{eq-decomp.105}). Hence ${\hat
\Gamma^{a}}_{\e}(p)\ge 0$. This proves positive definiteness of
$\Gamma^{a}_{\e}$. The positive definiteness of
$A^{a}_{\e}G^{a}_{\e}{A^{a}_{\e}}^{*} $ is obvious. 

Continuity of $\hat{\Gamma}^{a}_{\e} (p)$: By (\ref{eq-prel.6}),
\[
\hat{G}^{a}_{\e}(p) = \frac{1}{a-\hat{\D}_{\e_{n}} (p)} 
\] 
is continuous if $a>0$ or $p\not =0$. If $a=0$ then
(\ref{eq-decomp.165}) shows that the $p^{-2}$ singularity of
$\hat{\Gamma}^{a}_{\e}(p)$ is cancelled by
\[
\big|1-\hat{A}^{a}_{\e}(p)\big| 
=
\big|\hat{A}^{a}_{\e}(0)-\hat{A}^{a}_{\e}(p)\big|  = o(p^{2})
\]
which holds because $p$ derivatives of $\hat{A}^{a}_{\e}(p)$ are
moments for $A^{a}_{\e}$ and this is a  probability 
measure of compact support which therefore has moments of all orders.
\end{proof}

Choose $\e =\e_{n-1}$ for $n\ge 1$ and write (\ref{eq-decomp.12})  
in a rescaled form. It is easy to check that for $u,v\in (\e_{n-1}\bZ)^d
$ we have
\begin{equation}
G^{a}_{\e_{n-1}}(u-v)=L^{-(d-2)}G_{\e_{n}}^{L^{2}a}\big(\frac{u-v}{
L}\big) \label{eq-decomp.17}
\end{equation}  
\begin{equation}
\PP^{a}_{U_{\e_{n-1}}(R,z)}(x,du) = \PP^{L^{2}a}_{ U_{\e_{n}}(\frac{R}{
L}, \frac{z}{ L})}\big(\frac{x}{ L},\frac{du}{ L}\big) \label{eq-decomp.171}
\end{equation}

\noindent Define the sequence of functions  $g_{n}$ on ${\bf R}^{d}$ by 
\begin{equation}
g_{n}(z)= L^{nd}g(L^{n}z) \label{eq-decomp.20}
\end{equation}
where $g$ is the function introduced earlier (see
(\ref{eq-decomp.7a}), (\ref{eq-decomp.7b})) and observe that because
of the normalization (\ref{eq-decomp.7b}), the function $g_{n}$ is
also normalized :
\begin{equation}
\int_{{\bf R}^{d}}dz\ \ g_{n}(z)=1 \label{eq-decomp.21}
\end{equation}
\noindent Moreover, from the support property of $g$, we have that
\begin{equation}
g_{n}(x)=0\> :\> |x|\ge \frac{1}{ 4L^{n-1}} \label{eq-decomp.22}
\end{equation}
Let
\begin{equation}\label{eq-decomp.225}
R_{m}=L^{-(m-1)}
\end{equation} 
As in (\ref{eq-decomp.1021}) we have in $(\e_{n}\bZ)^d$ 
for $n\ge m\ge 0$  the measure 
$A^{a}_{\e_{n},m}(R_{m})(x,du)$ given by
\begin{equation}
\int_{(\e_{n}\bZ)^d}du\> A^{a}_{\e_{n},m}(R_{m})(x,du)\> f(u)
=\int_{(\e_{n}\bZ)^d}dz\>
c_{\e_{n-m}} g_{m}(x-z)
\PP^{a}_{U_{\e}(R_{m},z)}(x,f) \label{eq-decomp.23} \end{equation} 
Note that
\begin{equation}
A^{a}_{\e_{n},0}=A^{a}_{\e_{n}} \label{eq-decomp.23a}
\end{equation}
as defined earlier. Observe that 
\begin{equation}\label{eq-decomp.23b}
\int_{(\e_{n}\bZ)^d}dz\> c_{\e_{n-m}} g_{m}(z)= \int_{(\e_{n-m}\bZ)^d}dz\>
c_{\e_{n-m}}g(z)=1 
\end{equation}
from the definition of the constants $c_{\e_{n}}$ in
(\ref{eq-decomp.8}).  The definition (\ref{eq-decomp.20}) together
with (\ref{eq-decomp.23}) and  (\ref{eq-decomp.171}) imply the scaling relation
\begin{equation}
A^{a}_{\e_{n-1},m-1}(R_{m-1})(x,du) = A^{L^{2}a}_{\e_{n},m}(R_{m})
(\frac{x}{ L},\frac{du}{ L}) \label{eq-decomp.24}
\end{equation}
Applying this to the righthand side of (\ref{eq-decomp.12}) we get for
$n\ge 1$
\begin{equation}
G^{a}_{\e_{n-1}}(x-y) =\Gamma^{a}_{\e_{n-1}}(x-y) + L^{-(d-2)}{\bigl
(A^{L^{2}a}_{\e_{n},1}(1)\> G^{L^{2}a}_{\e_{n}}\>
A^{L^{2}a}_{\e_{n},1} (1)^{*}\bigr )} \bigl (\frac{x-y}{ L} \bigr )
\label{eq-decomp.26}
\end{equation} 
We can now iterate (\ref{eq-decomp.26}) starting with $n=1$, $n$-times
using the same principle. Define for $n\ge 1$
\begin{equation}
{\cal A}^{a}_{n}=\prod_{j=1}^{n}\> A^{a}_{\e_{n},n+1-j}(L^{-(n-j)})
\label{eq-decomp.28}
\end{equation}
For $n=0$ we set 
$${\cal A}^{a}_{0}=1$$
We also define on the $(\e_{n}\bZ)^d$ lattice
\begin{equation}
\Gamma^{a}_{n}={\cal A}^{a}_{n}\> \Gamma^{a}_{\e_{n}}\> {{\cal
A}^{a}_{n}}^{*} \label{eq-decomp.29}
\end{equation}
and  
\begin{equation}
\GG_{n}^{a} = {\cal A}^{a}_{n}\> G^{a}_{\e_{n}}\>
{{\cal A}^{a}_{n}}^{*} \label{eq-decomp.30}
\end{equation}
Then we have the \emph{multiscale decomposition for the resolvent}
\begin{equation}
G^{a}(x-y)=\sum_{j=0}^{n-1}\> L^{-j(d-2)}\> \Gamma_{j}^{L^{2j}a}{\bigl
(\frac{x-y}{ L^{j}}\bigr )}\> + L^{-n(d-2)}\> \GG^{L^{2n}a}_{n}\> \bigl
(\frac{x-y}{ L^{n}}\bigr ) \label{eq-decomp.31}
\end{equation} 
which is valid for $a\ge 0$. The special case $a=0$ gives the 
{\it multiscale decomposition for the massless Green's function}
\begin{equation}
G^{0}(x-y)=\sum_{j=0}^{n-1}\> L^{-j(d-2)}\> \Gamma_{j}^{0}{\bigl
(\frac{x-y}{ L^{j}}\bigr )}\> + L^{-n(d-2)}\>
\GG^{0}_{n}\> \bigl
(\frac{x-y}{ L^{n}}\bigr ) \label{eq-decomp.311}
\end{equation}

\begin{lemma}\label{lem-decomp.2}
For all $n\ge 0$
$$\Gamma_{n}^{a}(x-y)=0\> : \> |x-y|\ge 6L$$
\end{lemma}

\begin{proof} $\Gamma_{n}^{a}$ is a multiple convolution, so the range of
$\Gamma_{n}^{a}$ is the sum of the ranges of the convolved functions.
>From the definition (\ref{eq-decomp.23}) of $A^{a}_{\e_{n},m}(R)$, the
support property (\ref{eq-decomp.22}) of $g_{m}$ and
$\PP^{a}_{U_{\e_{n}}}(R)(x,u)$ vanishes if $|u-z| > R$, we find that
the range of $A^{a}_{\e_{n},m}(R)$ is $R+\frac{1}{ 4L^{m-1}}$. From
the definition (\ref{eq-decomp.28}) of $\AA^{a}_{n}$ the range of
$\AA^{a}_{n}$ is $\sum_{j=1}^{n}L^{-(n-j)} (1+\frac{1}{ 4L})$ which is
less than $3$ for $L$ large. By construction the range of 
$\Gamma^{a}_{\e_{n}}$ is less than $3L$  by  
Lemma~\ref{lem-decomp.1}. 
 The Lemma follows.
\end{proof}

\section{Probabilistic Aspects}\label{sec-probability}
\setcounter{equation}{0}


\vskip0.5cm \noindent 
\emph{Multiscale decomposition for the L\'evy Green's
function.}  
\vskip0.3cm
Let $x^{(\alpha)}_{t}$ , $0 < \alpha < 2$, be the  stable  
L\'evy process in $Z^d$, (stable in the sense that its scaling limit
is stable). Note that $E_{x}(|x^{(\alpha)}_{t}|) < \infty $, provided
$\alpha > 1$.  However in the following we will allow for the 
full admissable range
 $0 < \alpha < 2$.  The L\'evy Green's function $C$ is  
given by 
\begin{equation}
 C(x-y)=(-\Delta )^{-\alpha /2}(x-y)=const\int_{0}^{\infty}da\ \
a^{-\alpha /2}G^{a}(x-y) \label{eq-prob.8}
\end{equation}
with  $0 < \alpha < 2$. In this range it is easy to verify that the 
integral representation converges.  

$C$ has a finite range multiscale decomposition, which we
obtain by inserting the multiscale decomposition (\ref{eq-decomp.31})
for $G^{a}$ into the integral representation (\ref{eq-prob.8}) of the
L\'evy Green's function $C$ to get
\begin{gather*}
C(x-y)=\sum_{j=0}^{n-1}\> L^{-j(d-2)}\> const\int_{0}^{\infty}da\ \ 
a^{-\alpha /2}\Gamma_{j}^{L^{2j}a}{\bigl (\frac{x-y}{ L^{j}}\bigr )}\>
+ \nonumber\\
+ L^{-n(d-2)}\>  const\int_{0}^{\infty}da\ \ a^{-\alpha /2} 
\>  
\GG^{ L^{2n}a}_{n}\> 
\bigl (\frac{x-y}{ L^{n}}\bigr )
\end{gather*}
After rescaling in $a$ in each term we get
\begin{equation}
C(x-y)=\sum_{j=0}^{n-1}\> L^{-2j[\phi]}\> \Gamma_{j}{\bigl (\frac{x-y}{
L^{j}}\bigr )}\> + L^{-2n[\phi]}\> C_{n}\bigl (\frac{x-y}{ L^{n}} \bigr
) \label{eq-prob.32}
\end{equation} 
where
\begin{align}\label{eq-prob.325}
\Gamma_{j} & =const\int_{0}^{\infty}da\ \ a^{-\alpha /2}\> 
\Gamma_{j}^{a} \cr
C_{n} & =const\int_{0}^{\infty}da\ \ a^{-\alpha /2}\>
\GG_{n}^{a} \cr
[\phi] & =\frac{d-\alpha}{ 2} 
\end{align}
Note that $[\phi]$ as defined above is the canonical dimension of the
the scalar Gaussian field $\phi$ distributed with covariance $C$. We
have $[\phi]>0$, since $d\ge 3$ and $ 0< \alpha <2$.  $\Gamma_{j}$ and
$C_{n}$ are well defined because of the bounds provided below (see
section 5, Corollary~\ref{cor-bounds.6}).  Moreover by
Lemma~\ref{lem-decomp.2} and (\ref{eq-prob.325})
\[
\Gamma_{n}(x-y)=0\> :\> |x-y|\ge 6L 
\]
>From Lemma~\ref{lem-decomp.2} and (\ref{eq-prob.325}), $C_{n}$ and
$\Gamma_{n}$ are positive semi-definite and thus qualify as covariances
of Gaussian measures denoted $ \mu_{C_{n}}, \mu_{\Gamma_{n}}$.
The multiscale decomposition (\ref{eq-prob.32}) now gives rise to
renormalization group transformations. From (\ref{eq-prob.32}) we get
for $x,y\in (\e_{n}\bZ)^d $
\begin{equation}
C_{n}(x-y)=\Gamma_{n}(x-y)+L^{-2[\phi]}C_{n+1}(\frac{x-y}{ L})
\label{eq-prob.33}
\end{equation}
and hence we have a sequence of RG transformations
\begin{equation}
z_{n+1}(\phi)=\int d\mu_{\Gamma_{n}}(\zeta)\> z_{n}(\zeta
+\phi_{L^{-1}}) \label{eq-prob.34}
\end{equation}
where $$\phi_{L^{-1}}(x)=L^{-[\phi]}\phi(\frac{x}{ L})$$
and 
\begin{equation}
\int d\mu_{C_{n}}(\phi)\> z_{n}(\phi) =\int d\mu_{C_{n+1}}(\phi)\>
z_{n+1}(\phi)\label{eq-prob.35}
\end{equation} 
with $C_{0}=C$ given by (\ref{eq-prob.8}).

\vskip0.5cm \noindent \noindent
\emph{Poisson Kernel.} 
\vskip0.3cm

Let $x_{t}^{(n)},\> t\ge 0$ be continuous time simple random walk with
right continuous paths and state space $(\e_{n}\bZ)^{d}$. 
The characteristic function is
\[
E(e^{ip.x_{t}^{(n)}})=e^{t\hat{\D}_{\e_{n}}(p)}
\]
and the generator of the Markov process $x^{(n)}_{t}$ is the lattice
Laplacian $\Delta_{\e_{n}}$. Note that the semigroup
$e^{t\Delta_{\e_{n}}}$ is a contraction on
$L^{\infty}((\e_{n}\bZ)^d)$. In the discrete topology the latter
coincides with the space of bounded continuous functions. Hence
$e^{t\Delta_{\e_{n}}}$ is a Feller semigoup so that $x^{(n)}_{t}$ is
strong Markov with respect to stopping times $\tau$. 

Now let $x_{t}^{\e}$ be the above process in $(\e \bZ)^d$ . Let $P_{x}$
the probability measure for the process conditioned to start at $x$. As
in (\ref{eq-prel.22}) and (\ref{eq-prel.2}), $U(R)$ is an open cube of
radius $R$ in ${\bf R}^{d}$, $U_{\e}(R)$ the induced cube in $(\e
\bZ)^{d}$, and the boundary $\partial U_{\e}(R)$ is defined as in
(\ref{eq-prel.3}). Let $\tau_{U_{\e}}$ be the first exit time from
$U_{\e}$ .  Then $\tau_{U_{\e}}$ is also the first hitting time of
$\partial U_{\e}$ from the interior. $E_{x}(\tau_{U_{\e}}) < \infty$
because $U_{\e}(R)$ is bounded, and hence $\tau_{U_{\e}} <\infty,\>
P_{x}\> a.s.$.
  
Let $A$ be a subset of points in $\partial U_{\e}$. It is a standard
result in probability that the measure $\PP^{a}_{U_{\e}}(x,dy)$
defined on $\partial U_{\e}$ by
\begin{equation}
 \PP^{a}_{U_{\e}}(x,A) =
E_{x}(e^{-a\tau_{U_{\e}}}1_{x_{\tau_{U_{\e}}} \epsilon A})
\label{eq-prob.659}
\end{equation}
is the Poisson kernel we defined in (\ref{eq-decomp.660}).

Also, since $ P_{x}( x_{\tau_{U_{\e}}} \in \partial U)=1$ the total
mass is
\begin{equation} \label{eq-prob.337}
\PP^{a}_{U_{\e}}(x,\partial U_{\e}) =
E_{x}(e^{-a\tau_{U_{\e}}})\le 1 
\end{equation} 
The same construction works in ${\bf R}^{d}$ with $x^{\e}_{t}$ replaced 
by standard Brownian motion $x_{t}$, $U_{\e}(R)$ replaced by $U(R)$ and
$A$ taken to be any Borel subset of $\partial U(R)$.

Let now $R=R_{m}=L^{-(m-1)}$, with $0\le m \le n$. In Section 6 we will
need an estimate for $1-\PP^{a}_{U_{\e_{n}}}(x,\partial U_{\e_{n}})$
where it is understood $U_{\e_{n}}=U_{\e_{n}}(R_{m})$. Observe that
$1-e^{-a\tau}\le a \tau$ so
\[
0\le 1-\PP^{a}_{U_{\e_{n}}}(x,\partial U_{\e_{n}}) \le 
aE_{x}(\tau_{U_{\e}}).
\]
We can estimate the mean exit time as follows. Let $f$ be a smooth
function in ${\bf R}^{d}$ bounded in a neighbourhood of $U (R)$. Then
by the strong Markov property, for any $x\in U_{\e_{n}}(R_{m}) $,
\[
E_{x}\Bigl (\int_{0}^{\tau_{U_{\e}}}dt
(\Delta_{\e_{n}}f)(x_{t})\Bigr ) =E_{x}(f(x_{\tau_{U_{\e}}})) -f(x)
\]
We choose $x=0$ and $f (x) = |x|^{2}$ in $U(\frac{3}{2}R_{m})$. Then by
a simple computation in $U(R_{m})$ we obtain
$\Delta_{\e_{n}}f(x)=2d$. Moreover as shown in Section 6,
(\ref{eq-convergence.10}), $\partial U_{\e_{n}}(R_{m}) \subset
\partial U(R)$.  Hence we get the bound
\[
E_{x}(\tau_{U_{\e}})\le \frac{1}{2d} E_{x}(x^{2}_{\tau_{U_{\e}}}) \le
\frac{R_{m}^{2}}{ 2  } .
\]
Thus we have proved the following 
\begin{lemma}\label{lem-prob.1}
Let $R_{m}=L^{-(m-1)}, 0\le m\le n $ Then we have the bound
$$0\le 1-\PP^{a}_{U_{\e_{n}}(R_{m})}(x,\partial U_{\e_{n}}(R_{m}))
\le a \frac{R_{m}^{2}}{ 2 }$$
\end{lemma}

Now recall the definition of the measure $A_{\e_{n},m}^{a}(R_{m})(0,du)$
on $(\e_{n} \bZ)^{d}$ given in (\ref{eq-decomp.23}). $c_{\e_{n-m}} g_{m}(z)$
is a probability density in $(\e \bZ)^{d}$. Hence from 
 (Lemma~\ref{lem-prob.1}) we get 
\begin{corollary}\label{cor-prob.2}
$$0\le 1-\int_{(\e_{n} \bZ)^{d}}A_{\e_{n},m}^{a}(R_{m})(0,du) 
\le a \frac{R_{m}^{2}}{ 2 }$$
\end{corollary}

\section{Bounds}\label{sec-bounds}
\setcounter{equation}{0}

We will give uniform bounds on Fourier transforms and Sobolev norms of
arbitrary high index.  In Section~\ref{sec-convergence} we will prove
in the latter norms the convergence of the sequences $C_{n}$,
$\Gamma_{n}$. The limiting covariances will thus turn out to be in
$C^{p}$, $\forall p\ge 0$.

We recall the definition of the lattice derivative $\nabla_{\underline
e}$ in (\ref{eq-prel.4a}).   In particular
$\Scal=\{\hat{e}_{1},...\hat{e}_{d} \}$  is the standard
basis of unit vectors.  For $\underline e \in \Scal$,  $\nabla_{\underline
e}$ is the forward partial derivative and for
$-\underline e \in \Scal$,  $\nabla_{\underline
e}$ is the backward partial
derivative.  We define the $n$ th lattice derivative
\[
\nabla^{n}_{{\underline e_1},...,{\underline e_n}}=
\nabla_{\underline e_1}.....\nabla_{\underline e_n}
\]
Let $X$ be a connected open set in ${\bf R}^{d}$. We define
\[
X_{\e}= X \cap (\e \bZ)^d
\] 
We now define the lattice Sobolev norm \ \ $\Vert
.\Vert_{H_{k}(X_{\e})}$ of a function $f$ by
\begin{equation}
\Vert f\Vert^{2}_{H_{k}(X_{\e})}=\sum_{j=0}^{k} 2^{-j}
\sum_{{\pm\underline e_1},...,
{\pm\underline e_j}\in \Scal}
\int_{X_{\e}}dx \vert
\nabla^{j}_{{\underline e_1},...,{\underline e_j}}f(x)\vert^{2}
\label{eq-bounds.14}
\end{equation}

\begin{lemma}\label{lem-bounds.1a}
Let $g$ be a $C^{\infty}({\bf R}^{d})$ function.
Then for  every $k\ge 0$ there exists a 
constant $C_{k}$ independent of $\e$ such that 
\[
\Vert fg\Vert_{H_{k}(X_{\e})} \le 
C_{k}  \Vert g\Vert_{C^{k}({\bf R}^{d})} 
\Vert f\Vert_{H_{k}(X_{\e})}
\]
For every $k\ge 0$, and any $L_{1}$ function $g$,
\[
\Vert f\ast g\Vert_{H_{k}((\e \bZ)^{d}))} \le 
\Vert g\Vert_{L_{1} ((\e \bZ)^{d})}
\Vert f\Vert_{H_{k}((\e \bZ)^{d}))} 
\]
\end{lemma}

\begin{proof}
To prove this we take the square of the norm
on the left hand side and then  use the 
$(\e \bZ)^{d}$ lattice modification of the Leibniz rule :
\begin{equation}\label{eq-bounds.142}
\nabla_{\e,\underline e}(fg)=(\nabla_{\e,\underline e}f)g
+f\nabla_{\e,\underline e}g +\e\nabla_{\e,\underline e}f
\nabla_{\e,\underline e}g
\end{equation}

Derivatives on $g$ are bounded in the $C^{k}({\bf R}^{d})$ norm and  
all $\e$ dependent constants can be majorised by setting $\e =1$. This proves
the first inequality. The second inequality, which is a form of Young's 
convolution inequality, is proved exactly as in the continuum.
\end{proof}

In the following we will exploit a lattice version of elliptic
regularity.  Let $f$ be a bounded function in $(\e \bZ)^d$. Let
$U(R)\subset {\bf {\bf R}^{d}}$ be an open cube centered at the origin
and of edge length $R$.  Let $U_{\e}(R)=U(R)\cap (\e \bZ)^d $. Let $\O
\subset U(\frac{1}{4}R)$ be an open connected set. Then $\O_{\e}\subset
U_{\e}(\frac{1}{4}R)$.  Define
\[
h^{a}(x)=\PP^{a}_{U_{\e}}(x,f)
\]
Recall from Section 2 (see (\ref{eq-decomp.661}),
(\ref{eq-decomp.662})) et seq.) that $h^{a}(x)$ is the solution of the
Dirichlet problem
\[
(-\Delta_{\e}+ a)h^{a}(x)=0 : \ \ x\in U_{\e}  
\]
\begin{equation}
h^{a}(u)= f(u) :\ \ u\in \partial U_{\e}
\label{{eq-bounds.14.5}}
\end{equation}
and that the maximum principle holds, because $a\ge 0$. We have

\begin{proposition}[lattice elliptic regularity]\label{prop-bounds.4}
With $\O_{\e}$ defined as above, and $\forall \>k\ge 0$
\[
\Vert h^{a}\Vert_{H_{k}(\O_{\e})} \le C_{R} (1+a)^{-\frac{1}{2} } 
\Vert f\Vert_{L^{\infty}((\e \bZ)^{d})}.
\]
\end{proposition}

\noindent {\it Remark :}\ \ This is well known in the continuum.  For
completeness we give a proof of the lattice version in
Appendix~\ref{app-A}. 

We choose $\e = \e_{n}$ and apply the proposition to
\[
\big(A^{a}_{\e_{n},m=1} (L^{0})f\big)(x') 
:= \int_{(\e_{n}\bZ)^{d}}
\, dz \, c_{\e_{n-1}}g_{1} (x'-z)
\PP^{a_{n}}_{U_{\e_{n}} (1,z)} (x',f)
\]
which is the $j=n$ term in the product (\ref{eq-decomp.28}) defining
$\AA^{a}_{n}$.

\begin{corollary}\label{cor-bounds.4a} Let $\O_{\e_{n}}$ be as in
Proposition~\ref{prop-bounds.4} with $R=1$. Then for every $k\ge 0$ 
and every $n\ge 1$ there 
exists a constant $C_{k,L}$ independent of $\e_{n}$ such that  
\begin{equation}\label{eq-bounds.295}
\Vert 
A^{a}_{\e_{n},m=1} (L^{0})f
\Vert_{H_{k}(\O_{\e_{n}})}
\le 
C_{k,L}(1+a)^{-\frac{1}{2}}
\Vert
f\Vert_{L^{\infty}((\e_{n}\bZ)^{d})}. 
\end{equation}
\begin{equation}\label{eq-bounds.297}
\Vert 
\nabla^{j}_{{\underline e_1},...,{\underline e_k}}
A^{a}_{\e_{n},m=1}(L^{0})f
\Vert_{L_{\infty}((\e_{n}\bZ)^{d})}
\le 
C_{k,L} (1+a)^{-\frac{1}{2}}
\Vert
f\Vert_{L^{\infty}((\e_{n}\bZ)^{d})}.
\end{equation}
\end{corollary}

\begin{proof} The integral over $z$ in
(\ref{eq-bounds.295},\ref{eq-decomp.10}) can be restricted
to $U_{\e_{n}} (2)$ because the range of $g_{1}$ is
$1/4$. Therefore
\begin{gather*}
\Vert 
A^{a}_{\e_{n},m=1} (L^{0})(\cdot,f)
\Vert_{H_{k}(\O_{\e_{n}})}
\le \\
\int_{U_{\e_{n}} (2)}
\, dz \, c_{\e_{n-1}}\Vert g_{1}\Vert_{C^{k}( {\bf R}^{d})}
\Vert\PP^{a_{n}}_{U_{\e_{n}} (1,z)} (\cdot,f)\Vert_{H_{k}(\O_{\e_{n}})}\\
\le
c_{k,L}(1+a)^{-\frac{1}{2}} 
\Vert f\Vert_{L^{\infty}((\e_{n}\bZ)^{d})} 
\end{gather*}
We have used the first inequality of Lemma~\ref{lem-bounds.1a}, absorbing
the $C^{k}({\bf R}^{d})$ norm of $g_{1}$ in the constant since $g$ has been
fixed once for all , and then
using Proposition~\ref{prop-bounds.4}.

To prove (\ref{eq-bounds.297}): By the embedding of high degree
Sobolev space into $L_{\infty}$, (Lemma~\ref{lem-B.1}), reviewed in
Appendix~\ref{app-B}, we pass from (\ref{eq-bounds.295}) to
\begin{equation}\label{eq-bounds.296}
\Vert 
\nabla^{j}_{{\underline e_1},...,{\underline e_k}}
A^{a}_{\e_{n},m=1} (L^{0})f
\Vert_{L_{\infty}((\e_{n}\bZ)^{d})}
\le 
 C_{k,L} (1+a)^{-\frac{1}{2}}
\Vert f\Vert_{L^{\infty}((\e_{n}\bZ)^{d})}.
\end{equation}
noting that (\ref{eq-bounds.295}) applies to any translate
$x+\O_{\e_{n}}$ of $\O_{\e_{n}}$.  
\end{proof}

\begin{lemma}\label{lem-bounds.7}
For every integer $k\ \ge \ 1$,  and every $n\ge 1$,  
$\exists$ a constant $c_{k,L}$ independent of $\e_{n}$ such that ,
\begin{equation}
\big\vert \hat{A}_{\e_{n},m=1}^{a}(L^{0})(p) \big\vert \le
c_{k,L} (1+a)^{-\frac{1}{2}}
\bigl (-{\hat {\D}}_{\e_{n}}(p) + 1 \bigr )^{-k}
\label{eq-bounds.37} 
\end{equation}
\end{lemma}

\begin{proof} This is the essentially the standard proof that the
Fourier transform of a smooth function of compact support has rapid decay. Let
\[
h (x,p) = \int_{(\e_{n}\bZ)^{n}}A_{e_{n},m=1}^{a}
(L^{0})(x,du)e^{-ipu} 
\]
Then
\[
\bigl (-{\hat {\D}}_{\e_{n}}(p) + 1 \bigr )^{k}h (x,p) =
\int_{(\e_{n}\bZ)^{n}} A_{e_{n},m=1}^{a}(L^{0})(x,du)
\bigl (-\D_{u,\e_{n}} + 1 \bigr )^{k}e^{-ipu}
\]
where the $u$ subscript on $\D_{u,\e_{n}}$ indicates the variable it
differentiates. Since $\D_{u,\e_{n}}$ is a linear combination of
lattice translations under which the lattice is invariant, the exact
analogue of integration by parts is valid and we continue with
\[
 = \int_{(\e_{n}\bZ)^{n}} \bigl (-\D_{u,\e_{n}} + 1 \bigr
)^{k} A_{\e_{n},m=1}^{a}(L^{0})(x,du) e^{-ipu}
\]
By the translation invariance (see (\ref{eq-decomp.102})) of
$A_{\e_{n},m=1}^{a}(L^{0})(x,du)$ we can change the derivatives to $x$
\[
 = \int_{(\e_{n}\bZ)^{n}} \bigl (-\D_{x,\e_{n}} + 1 \bigr)^{k} 
A_{\e_{n},m=1}^{a}(L^{0})(x,du) e^{-ipu}
\]
By (\ref{eq-bounds.297}) in Corollary~\ref{cor-bounds.4a} with $f
(u)=\exp (ip.u)$,
\[
\bigg|
\int_{(\e_{n}\bZ)^{n}} \bigl (-\D_{x,\e_{n}} + 1 \bigr
)^{k} A_{\e_{n},m=1}^{a}(L^{0})(x,du) e^{-ipu}
\bigg|
\le
c_{k,L}(1+a)^{-\frac{1}{2}} 
\]
Collecting these relations we have
\[
\bigg|
\bigl ({-\hat {\D}}_{\e_{n}}(p) + 1 \bigr )^{k}h (x,p)
\bigg| \le
c_{k,L}(1+a)^{-\frac{1}{2}} 
\]
By setting $x=0$ we finish the proof.
\end{proof}

Fourier transforms are naturally defined on the Brillouin zone
\begin{equation}\label{eq-bounds.293}
B_{\e}=[-\pi
/\e,\pi/\e]^{d}
\end{equation}
There is a constant $c$ independent of $\e$ such that
\begin{equation}\label{eq-bounds.294}
p^{2} \ge -\hat{\D}_{\e} (p) \ge cp^{2} \text{ for } p \in B_{\e}
\end{equation}
which follows from $t^{2}/2 \ge 1-\cos t \ge ct^{2}$ on $[-\pi ,\pi]$.

\begin{theorem}\label{thm-bounds.5}
${\forall n}\ge 0$ and $\forall k\ge 0$, 
$\exists$ a constant $ c_{k,L}$
independent of $n$  such that
\begin{equation}
|\hat {\Gamma}^{a}_{n}(p)| \le  c_{k,L}
(1+a)^{-1}
(1+p^{2})^{-2k} \text{ for }
p \in B_{\e_{n}}
\label{eq-bounds.391}
\end{equation}
\begin{equation}
\Vert \G^{a}_{n}\Vert_{H_{k}((\e_{n}\bZ)^{d})} \le 
c_{k,L}(1+a)^{- 1} 
\label{eq-bounds.29}
\end{equation}

\end{theorem}

\begin{proof} 
>From (\ref{eq-prel.6}) and (\ref{eq-bounds.294}),
\[
0\le \hat{G}^{a}_{\e_{n}} (p) \le (-\hat{\D}_{\e_{n}})^{-1} 
\le \frac{c}{p^{2}}
\]
Combining this with (\ref{eq-decomp.165}) and the
continuity assertion of Lemma~\ref{lem-decomp.1} we have
\[
\hat{\Gamma}^{a}_{\e_{n}}(p) \le c_{L} (1+p^{2})^{-1}
\]
\noindent \emph{Case} $n=0$:
It is sufficient to prove that $\hat{\Gamma}^{a}_{\e_{0}}(p)$ is
bounded  by $C (1+a)^{-1}$ with $C$ uniform in $p$ because
$B_{\epsilon_{0}}$ is bounded. Referring to
(\ref{eq-decomp.165},\ref{eq-prel.6}) we find that
\[
        |\hat{\Gamma}^{a}_{\e_{0}}(p)|
\le 
        2\frac{|1-\hat{A}^{a}_{\e_{0}}(p)|}{a-\hat{\D}_{\e_{0}}(p)}
\le
        2\frac{|1-\hat{A}^{a}_{\e_{0}}(0)|}{a} + 
        2\frac{|\hat{A}^{a}_{\e_{0}}(0)-\hat{A}^{a}_{\e_{0}}(p)|}
        {a-\hat{\D}_{\e_{0}}(p)}
\]
The first term is continuous at $a=0$ by
Corollary~\ref{cor-prob.2}. Therefore it is bounded by $C
(1+a)^{-1}$. The second term is bounded by $C (1+a)^{-1}$ using the
same argument (existence of moments of the Poissson measure) as in the
proof of Lemma~\ref{lem-decomp.1}.

\noindent \emph{Case} $n\ge 1$:    
by Lemma~\ref{lem-bounds.7}, and the bound $|{\hat
A}^{a}_{\e_{n},m}(p)|\le 1$ which we use for $m\ge 2$
\[
|\hat{\AA}^{a}_{n}(p)|^{2}\le 
C_{k,L} (1+a)^{-1} 
\bigl (p^{2} + 1 \bigr )^{-k} \label{eq-bounds.38}
\]
Using these estimates in (\ref{eq-decomp.29}) we obtain
(\ref{eq-bounds.391}).

Proof of (\ref{eq-bounds.29}). We have the easily established
bound
$$| p^{2} +{\hat \D}_{\e_{n}}(p)| \le O(1)\e_{n}^{2}|p|^{4}$$
so that
$$ 0\le 1-{\hat \D}_{\e_{n}}(p)\le (1+p^{2})(1+O(1)\e_{n}^{2}p^{2})$$
and hence for any $m\ge 0$
$$(1-{\hat \D}_{\e_{n}}(p))^{m}|{\hat {\Gamma}^{a}_{n}(p)}|^{2}
\le O(1)(1+p^{2})^{2m}|{\hat {\Gamma}^{a}_{n}(p)}|^{2}
\le c_{k,q,L} (1+a)^{-1} 
(1+p^{2})^{-q}$$
for any $q \ge 0$ by choosing $k$ in (\ref{eq-bounds.391}) sufficiently large.
Taking $q > d$  proves the theorem  because
$$\Vert \G^{a}_{n}\Vert_{H_{m}((\e_{n}\bZ)^{d})}^{2}
\le \int_{[-\frac{\pi}{\e_{n}} ,\frac{\pi}{\e_{n}}]^{d}}
\frac{d^{d}p}{{(2\pi)}^{d}} 
(1-{\hat \D}_{\e_{n}}(p))^{m}| {\hat {\Gamma}^{a}_{n}(p)|}^{2}$$

\end{proof}

\vskip0.3cm  Now turn to the L\'evy fluctuation
covariance given in (\ref{eq-prob.325}). Using
the bounds provided in Theorem~\ref{thm-bounds.5} we get 

\begin{corollary}\label{cor-bounds.6}
For $0< \a < 2$, all $k=0,1,\dotsc $,  and all $n\ge 0$,
\begin{equation}
\Vert \G_{n}\Vert_{H_{k}((\e_{n}\bZ)^{d})} \le c_{ k  ,L} 
\label{eq-bounds.31}
\end{equation}
where the constant on the right hand side is independent of $n$.
\end{corollary}

\section{Convergence}\label{sec-convergence}
\setcounter{equation}{0}

Theorem~\ref{thm-bounds.5} and Corollary~\ref{cor-bounds.6} provide
uniform bounds in Sobolev norms for fluctuation and block
covariances. In particular they are uniform in the lattice spacing
$\e_{n}$. We will now prove that these sequences converge to their
formal continuum limits.  Continuum objects have the subscript $c$ in
place of $\e$. Thus, as in (\ref{eq-prel.5}),
\begin{gather} 
G_{c}^{a} (x-y) = \int_{\bR^{d}}\frac{d^{d}p}{ (2\pi)^{d}}\>
e^{ip.(x-y)} \hat{G}_{c}^{a} (p) \label{eq-convergence.100}\\
\hat G^{a}_{c} (p) 
= (a + p^{2})^{-1}\label{eq-convergence.101}
\end{gather}
Recall from Section~\ref{sec-decomp} that
\[
U(R)\equiv U_{c} (R) =\bigl (-\frac{R}{2},\frac{R}{2}\bigr )^{d}\subset
\bR^{d}
\] 
represents an open cube of edge length $R$. In analogy to
(\ref{eq-decomp.23}) with $c_{c}:=1$ we define the continuum average
$A^{a}_{c,m}(R_{m})(x,,du)$ by
\begin{equation}
\int_{\bR^d}du\> A^{a}_{c,m}(R_{m})(x,du)\> f(u)
=\int_{\bR^d}dz\>
c_{c}g_{m}(x-z)
\PP^{a}_{c,U(R_{m},z)}(x,f) \label{eq-convergence.25} 
\end{equation} 
where 
\[
\PP^{a}_{U_{c}(R,z)}(x,f)=\int_{\partial U_{c}(R,z)} du\  
\PP ^{a}_{U_{c}(R, z)}(x,u)f(u)\label{eq-convergence.13}
\] 
is the solution $h^{a}_{c}(x)$ to the continuum Dirichlet problem
\[
(-\Delta_{c}+ a)h^{a}_{c}(x)=0 : \ \ x\in U_{c}({R})
\label{eq-convergence.11}  
\]
\begin{equation}
h^{a}_{c}(x)= f(x) :\ \ x\in \partial U_{c}(R)\label{eq-convergence.12}
\end{equation}
With these notations, the Fourier transform of the continuum analogue
of (\ref{eq-decomp.12}) is
\begin{equation}
\hat{\G}^{a}_{c} (p)=G^{a}_{c}(p)-|{\hat A}^{a}_{c,0}(R_{0})(p)|^{2}
\hat{G}^{a}_{c} (p)\label{eq-convergence.80}
\end{equation}
and that of (\ref{eq-decomp.29}) is
\begin{equation}
\hat{\G}^{a}_{c,n}=\prod_{m=1}^{n} |{\hat A}^{a}_{c,m}(R_{m})|^{2}
\hat{\G}^{a}_{c} \label{eq-convergence.81}
\end{equation}
The Lemmas, Propositions, Theorems and their Corollaries of sections
3,4 and 5 remain true in the continuum with the following 
caveat : in the continuum  the uniform Sobolev bounds of Theorem 
\ref{thm-bounds.5} and Corollary \ref{cor-bounds.6} hold only for $n\ge 1$. 
Aside from this caveat their proofs are identical and
need no repetition.  When referring to them for the continuum objects
we shall simply mention them as the continuum analogues of the
relevant results for the lattice.

Recall that $B_{\e_{n}}$ is the Brillouin zone, defined in
(\ref{eq-bounds.293}). The main result is

\begin{theorem}\label{thm-convergence.1} 
For every integer $k\ge 0$, 
\begin{equation}
{\G}^{a}_{c,n} \rightarrow {\G}^{a}_{c,*}\label{eq-convergence.82}
\end{equation}
in $H_{k}({\bf R}^{d})$. Moreover, for every fixed lattice
$(\e_{l}\bZ)^{d}$, $0\le l\le n$ the restriction of $\G^{a}_{n} (x)$ to
$(\e_{l}\bZ)^{d}$ converges to the continuum $\G^{a}_{c,*} (x)$
restricted to $(\e_{l}\bZ)^{d}$  in the Sobolev norm
\begin{equation}
\Vert \G^{a}_{c,*} - \G^{a}_{n}\Vert_{H_{k}((\e_{l}\bZ)^{d})} 
\rightarrow 0 \text{ as } n\rightarrow \infty  \label{eq-convergence.2}
\end{equation}
Moreover  multiple lattice derivatives of $\G^{a}_{n}$ converge to
the corresponding continuum derivatives of $\G^{a}_{c,*}$  
in the $L_{\infty}((\e_{l}\bZ)^{d})$ norm.
\end{theorem}

For the Levy finite range decomposition (\ref{eq-prob.32}) we
apply the last theorem to (\ref{eq-prob.325}) and obtain  

\begin{corollary}\label{cor-convergence.2}  For all $k\ge 0$,
\begin{align}
\Vert \G_{c,*}- \G_{n} \Vert_{H_{k}((\e_{ l}\bZ)^{d})} 
\rightarrow 0 \text{ as } n \rightarrow \infty  \label{eq-convergence.3}
\end{align}
Moreover multiple lattice derivatives of
$\G_{n}$ converge to the corresponding continuum derivatives of $ \G_{c,*}$
in the $L_{\infty}((\e_{ l}\bZ)^{d})$ norm.
\end{corollary}

We now give some Lemmas which will be employed in the proof of
Theorem~\ref{thm-convergence.1}. In the following lemmas we consider
continuum functions $f:\bR^{d}\rightarrow \bR$ and use the same symbol for
the lattice function $f:\bZ^{d}\rightarrow \bR$ defined by
restriction. Continuum and lattice integration are to be distinguished
by the domain of integration.

\begin{lemma}\label{lem-convergence.2}
\[\Vert  f(.+h)-f(.)\Vert_{H_{k}(\bR^{d})}\le c\vert h\vert .
\Vert f\Vert_{H_{k+1}(\bR^{d})}\label{eq-convergence.5} \] where $\vert
h\vert$ is the norm in $\bZ^{d}\subset \bR^{d}$. Moreover if $\vert h\vert
\le R$ we have
\begin{equation}
\Vert  f(.+h)-f(.)\Vert_{H_{k}(U_{c}(R))}
\le c\vert h\vert .\Vert f\Vert_{H_{k+1}(U_{c}(2R))} \label{eq-convergence.6}
\end{equation}
\end{lemma}

\begin{proof} See Theorem 3.3  on page 42 of \cite{Agmon65}.

\end{proof}

\begin{lemma}\label{lem-convergence.3}
Define
\[Q = -\D_{\e} - (-\D_{c})  \label{eq-convergence.7}\]
Then we have for every $k\ge 0$,
\begin{equation}
\Vert Qf \Vert_{H_{k}(\bR^{d})}\  \le \ c \e \  \Vert f
\Vert_{H_{k+3}(\bR^{d})} \label{eq-convergence.8} 
\end{equation}
where the constant $c$  is independent of $\e$.
\end{lemma}

\begin{proof} $\nabla_{{\underline e},\e}$ is the forward lattice
derivative in $(\e \bZ)^{d}$ , and $\nabla_{{\underline
e},\e}^{*}$ , its $L^{2}((\e \bZ)^{d})$ adjoint, the backward
lattice derivative. Forward and backward derivatives commute.
$\nabla_{{\underline e},c}$ is the continuum derivative in direction
${\underline e}$ and the adjoint is $\nabla_{{\underline e},c}^{*} = -
\nabla_{{\underline e},c}$.  A calculation shows that
\begin{equation}\label{801}
\nabla_{{\underline e},\e}^{*}
\nabla_{{\underline e},\e}
f (x)
= 
\int_{0}^{1}dt\ \int_{0}^{1}ds\ 
(-\nabla_{{\underline e},c}^{2})
f \big(x+[t-s] \e{\underline e}\big)
\end{equation}
Therefore
\[
Qf (x) = -\sum_{\underline e} \int_{0}^{1}dt\ \int_{0}^{1}ds\ 
\bigg(\nabla_{{\underline e},c}^{2}
f \big(x+[t-s] \e{\underline e}\big) - 
\nabla_{{\underline e},c}^{2}
f (x)\bigg)
\]
since 
\[
-\D_{\e} = \sum_{{\underline e}} 
\nabla_{{\underline e},\e}^{*} 
\nabla_{{\underline e},\e}
\]
the lemma is proved by taking norms under the integrals and using
Lemma~\ref{lem-convergence.2} (which gives the factor $\e$).

\end{proof}

We can now describe the main idea.  Let $f:\partial U_{c}
(R)\rightarrow \bR$ be a continuum function $f:\bR^{d}\rightarrow \bR$
restricted to the continuum boundary of $U_{c} (R)$. We need to
estimate the difference between the solution $h^{a}_{c}$ to the
continuum Dirichlet problem and the solution $h^{a}_{\e_{n}}$ to the
lattice Dirichlet problem.  This will be done by restricting
$h^{a}_{c}$ to the lattice.  The restriction also solves a
\emph{lattice} Dirichlet problem, but with a non-zero right hand side
involving $Qh^{a}_{c}$, which by Lemma~\ref{lem-convergence.3} is
$O(\e_{n})$.  Thus we consider
\begin{equation}
h^{a}_{\e_{n}}(x)- h^{a}_{ c }(x)
\ :\ x\in U_{\e_{n}}(\frac{R}{4}) 
\subset U_{ c }(\frac{R}{4})
\end{equation}
\[h^{a}_{\e_{n}}(x)- h^{a}_{ c  }(x)=0
\ :\ x\in {\partial}U_{\e_{n}}(R)\]
in Sobolev norms. The difference satisfies zero boundary conditions
because we will arrange that the lattice boundary points $\partial
U_{\e_{n}} (R)$ all lie on the continuum boundary $\partial U_{c} (R)$
and both solutions have boundary values $f$ restricted to $\partial
U_{\e_{n}} (R)$.

The lattice cube 
\[
U_{\e}(R) = U(R)\cap (\e \bZ)^{d} 
\]
has as its boundary
\[
\partial U_{\e}(R) = \{y \not \in \> U_{\e}(R) :\>  |x-y|=\e,\> 
{\rm some} \> x\in U_{\e}(R)\} 
\] 
>From Section~\ref{sec-preliminaries} we have $\e_{n} = L^{-n},\
L= 2^{p},\ p\ge 1 $. In addition we now choose , in accord with
(\ref{eq-decomp.225}),$R=R_{m}=L^{-(m-1)}:\ 0\le m\le n$. Then not
only do we have
\[U_{\e_{n}}(R)\subset U_{c}(R):=U (R) \label{eq-convergence.9}\]
but also
\begin{equation}{
\partial U}_{\e_{n}}(R)\subset \partial U_{c}(R) \label{eq-convergence.10}
\end{equation}
This last statement follows from the observation that $\e_{n}=2^{-np}$
and $\frac{R_{m}}{2}=2^{-mp + p-1}$ so that for $0\le m \le n $ we have
$\frac{R_{m}}{2}\in \e_{n}\bZ \subset \bR$. This means that for $d=1$ the
boundary points of $ U_{\e_{n}}(R_{m})$ coincide with the boundary
points of $ U(R_{m})$. For $d> 1$ we are in cubes and the above
reasoning easily generalises to (\ref{eq-convergence.10}).

\begin{lemma}\label{lem-convergence.4}
Let
\[f\ : \ \bR^{d}\rightarrow \bR, \]
\noindent Then for
\[
R = L^{-(m-1)}\ :\ 0\le m \le n\]
we have
\[
\Vert 
\nabla^{k}_{{\underline e_1},...,{\underline e_k}} 
\big(h^{a}_{\e_{n}}- h^{a}_{c}\big)
\Vert_{L_{\infty}(U_{\e_{n}}(\frac{R}{4}))}
\le c_{L,R, k } \e_{n}\Vert f\Vert_{
L^{\infty}(\partial U_{c}(R)) }
\] 
where the constant $c_{L,R, k}$ is independent of
$n$.
\end{lemma}

\proof See Appendix A. 

To proceed further we need a formula. First define a new finite
difference derivative that acts on continuum functions by \ : 
\[
\tilde{\nabla}_{\underline{e},\e}f(z)
=  \int_{0}^{1}dt \ (1-t) \nabla_{\underline{e},c}f (z+t\e)
\label{eq-convergence.14}
\] 
Then we have

\begin{lemma}\label{lem-convergence.5}
\begin{equation}
\int_{\bR^{d}}dz\ f(z)-\int_{(\e \bZ)^{d}}dz\ f(z)=
\e \sum_{j=0}^{d-1}
\int_{(\e \bZ)^{d-j}\times \bR^{j}} dz\
\tilde{\nabla}_{\underline{e}_{j+1},c}f(z) \label{eq-convergence.15} 
\end{equation}
\end{lemma}

\begin{proof}
We obtain case $d=1$ by
\begin{gather*}
\int_{\bR} dz\ f(z) - \int_{\e \bZ} dz\ f (z) 
=
\sum_{z\in \e \bZ} \int_{[z,z+\e]}dx\ f (x)  - \int_{\e \bZ} dz\ f (z) \\
=
\int_{\e \bZ} dz\ \e^{-1}\int_{[z,z+\e]}dx\ 
\big(f (x) - f (z)\big)
=
\e\int_{\e \bZ} dz\ 
\tilde{\nabla}_{\underline{e},\e}f(z)
\end{gather*}
and then the general case is obtained by applying this formula
iteratively. 
\end{proof}

The Fourier transform of $ A^{a}_{\e_{n},m}(R_{m})(0,du)$ is given by
\[
{\hat A}^{a}_{\e_{n},m}(R_{m})(p)=c_{\e_{n-m}}\int_{(\e_{n}\bZ)^{d}} dz
\ g_{m}(z)e^{-ip.z}h^{a}_{\e_{n},m}(z,p) 
\] 
where
\[
h^{a}_{\e_{n},m}(z,p)=\int_{\partial U_{\e_{n}}(R_{m}} 
{\cal P}^{a}_{U_{\e_{n}}(R_{m})}(z,du)e^{ip.u} 
\] 
Likewise, there is the continuum Fourier transform ${\hat
A}^{a}_{c,m}(R_{m})(p)$ defined by the same formulas with $\e_n,
\e_{n-m}$ replaced by $c$ and with $c_{c}:=1$.

We wish to estimate the difference ${\hat
A}^{a}_{c,m+1}(R_{m+1})(p) - {\hat
A}^{a}_{\e_{n},m+1}(R_{m+1})(p)$.  This is provided by the following
Lemma :

\begin{lemma}\label{lem-convergence.6}
For all integers $k\ge 0$, and  $n\ge 1$,
$\exists$ a constant $c_{k,L,m}$ independent
of $n$  such that
\[\bigl\vert {\hat A}^{a}_{c,m+1}(R_{m+1})(p) -
{\hat A}^{a}_{\e_{n},m+1}(R_{m+1})(p)\bigr\vert \le c_{k,L,m}\ \e_{n}
\label{eq-convergence.16}\]
\end{lemma}

\begin{proof} It is easy to see using Lemma~\ref{lem-convergence.5}
that we can write
\begin{equation}{
\hat A}^{a}_{c,m+1}(R_{m+1})(p) -{\hat A}^{a}_{\e_{n},m+1}(R_{m+1})(p)
= R_{1}(p) + R_{2}(p) + R_{3}(p) \label{eq-convergence.17} 
\end{equation}
where 
\[
R_{1}(p)=(1-c_{\e_{n-1-m}}) \int_{(\e_{n}\bZ)^{d}}dz\ e^{-ip.z}
g_{m+1}(z)h^{a}_{c,m+1}(z,p) 
\]
\[
R_{2}(p)=c_{\e_{n-1-m}} \int_{(\e_{n}\bZ)^{d}}dz\ e^{-ip.z} g_{m+1}(z)\bigl (h^{a}_{c,m+1}(z,p)-
h^{a}_{\e_{n},m+1}(z,p)\bigr )
\]
\[
R_{3}(p)=\e_{n}
\sum_{j=0}^{d-1}\int_{(\e_{n}\bZ)^{d-j}\times {\bf R}^{j}}
dz\
e^{-ip.z} \tilde{\nabla}_{\underline e_{j+1},c}\bigl (
g_{m+1}(z)h^{a}_{c,m+1}(z,p)\bigr )
\]
We observe that
\[
|1-c_{\e_{n-1-m}}| \le c_{L}\e_{n}
\] 
as follows from the definition (\ref{eq-decomp.8}),
Lemma~\ref{lem-convergence.5}, and the fact that $g$ is a smooth
function in $\bR^{d}$ of compact support. The integral is bounded by
$O(1)$.

For the term involving $ R_{2}$ we use Lemma~\ref{lem-convergence.4}
with $f (u)=\exp (ip.u)$ which produces the small factor $\e_{n}$in
the bound.

Finally in $R_{3}$ the $O(\e_{n})$ factor is already there. In the
integrand $h^{a}_{c,m+1}(z,p)$ and derivatives are $L_{\infty}$
bounded on the support of $g_{m+1}$ by (the continuum versions of)
Proposition~\ref{prop-bounds.4} and Lemma~\ref{lem-B.1}, with $f
(u)=\exp (ip.u)$.
 
\end{proof}

\begin{proof} (Theorem~\ref{thm-convergence.1}) Let $\epsilon >0$ and
fix any $p\not =0$.  Fix any $a\ge 0$. We will first prove that
\[
\vert \hat{\G}^{a}_{c,*} (p) - \hat{\G}^{a}_{n} (p) \vert <\epsilon 
\]
for all sufficiently large $n$.

Recall that $\hat{A}^{a}_{\e_{n},m} (p)$ and $\hat{A}^{a}_{\e_{n},m}
(p)$ are Fourier transforms of defective probability measures
supported in a cube of side $R_{m}$. Now
$$
\vert 1 - \hat{A}^{a}_{\e_{n},m} (p) \vert \le
\Bigl \vert 1-\int {A}^{a}_{\e_{n},m}(0,du)\Bigr \vert
+ \int {A}^{a}_{\e_{n},m}(0,du)|1-\exp (-ip.x)|.
$$ 
Therefore, by $|1-\exp (-ip.x)| \le c|p.x|$, and
Corollary~\ref{cor-prob.2},
\begin{equation}
\vert 1 - \hat{A}^{a}_{\e_{n},m} (p)\vert
\le 
c  R_{m} \vert p \vert + c a R_{m}^{2} 
\label{eq-convergence.83}
\end{equation}
Note that the same bound holds in the continuum, because 
Corollary~\ref{cor-prob.2} remains true in the continuum.

>From the definition of ${\hat \G}^{a}_{c,n}$ we get 
$${\hat \G}^{a}_{c,n+1}-{\hat \G}^{a}_{c,n}=
\Bigl ( |{\hat A}^{a}_{c,n+1}(R_{n+1})|^{2}-1\Bigr ){\hat \G}^{a}_{c,n}$$
whence, using the continuum analogue of (\ref{eq-convergence.83}), 
$$|{\hat \G}^{a}_{c,n+1}-{\hat \G}^{a}_{c,n}|
\le  \Bigl (2c R_{n+1} \vert p \vert  + 2c a R_{n+1}^{2}\Bigr )
|{\hat \G}^{a}_{c,n}| $$
Now ${\hat \G}^{a}_{c,n}$ satifies the uniform bound of 
Theorem~\ref{thm-bounds.5}, $R_{n+1}$ decreases geometrically with increasing
$n$. Therefore we see from the previous inequality that
${\hat \G}^{a}_{c,n}$ form a Cauchy sequence. This proves the existence
of the limit ${\hat \G}^{a}_{c,*}$ satisfying the bound of 
Theorem~\ref{thm-bounds.5} and the first part of 
Theorem~\ref{thm-convergence.1} has been proved.

Choose $n$ sufficiently large so that
\begin{equation}
|\hat {\G}^{a}_{c,*}(p)-\hat {\G}^{a}_{c,n}(p)| < \frac{\e}{4}
\label{eq-convergence.84}
\end{equation} 
${\hat \G}^{a}_{\e_{n}}(p)$ is uniformly continuous in $p$ 
by Lemma~\ref{lem-decomp.1}. Therefore, as shown in the course of proving
Theorem~\ref{thm-bounds.5} , there exists
a constant $\gamma$  independent of $\e_{n}$ such that
\begin{equation}
|\hat {\G}^{a}_{\e_{n}}(p)|\le \gamma
\label{eq-convergence.85}
\end{equation}

\noindent Now (\ref{eq-convergence.83}) implies that
\[
\big\vert 
\vert \hat{A}^{a}_{\e_{n},m}\vert - 
\vert \hat{A}^{a}_{c,m}\vert
\big\vert 
\le
2c R_{m}\vert p\vert + 2 c a R_{m}^{2} 
\]
Since $R_{m}\rightarrow 0$ geometrically fast and
$|\hat{A}^{a}_{\ast,m} (p)|\le 1$, we can choose $N$ depending on
$\epsilon$ such that for all $n>N$,
\begin{equation}\label{eq-convergence.120}
\big\vert
\prod_{m>N}^{n}\vert \hat{A}^{a}_{c,m}\vert^{2} - 
\prod_{m>N}^{n}\vert \hat{A}^{a}_{\e_{n},m}\vert^{2}
\big\vert < \frac{\epsilon}{4\gamma}
\end{equation}
By Lemma~\ref{lem-convergence.6} there is a constant $C_{N}$ such that
\begin{equation}\label{eq-converge.121}
\big\vert
\prod_{m=1}^{N}\vert \hat{A}^{a}_{c,m}\vert^{2} - 
\prod_{m=1}^{N}\vert \hat{A}^{a}_{\e_{n},m}\vert^{2}
\big\vert 
<
C_{N}\e_{n} < \frac{\epsilon}{4\gamma} 
\end{equation}
for all sufficiently large $n$. Finally,
\begin{equation}\label{eq-converge.122}
\vert 
\hat{\Gamma}^{a}_{ \e_{n} } - 
\hat{\Gamma}^{a}_{c} 
\vert < \frac{\epsilon}{4}
\end{equation}
by the definitions (\ref{eq-decomp.12}) , the explicit Fourier
transforms \ref{eq-prel.6} \ref{eq-convergence.101}) and 
Lemma~\ref{lem-convergence.6}.

>From the definition (\ref{eq-decomp.29}) we see that the four
inequalities (\ref{eq-convergence.85}, \ref{eq-convergence.120},
\ref{eq-converge.121}, \ref{eq-converge.122}) imply that
\begin{equation}\label{eq-convergence.123}
|\hat {\G}^{a}_{c,n}-\hat {\G}^{a}_{n}| < \frac{3\epsilon}{4}
\end{equation}
(\ref{eq-convergence.123}) and (\ref{eq-convergence.84}) establish the
pointwise convergence
\begin{equation}\label{eq-converge.124}
\hat {\G}^{a}_{n}(p)\rightarrow \hat {\G}^{a}_{c,*}(p)
\end{equation}

By the dominated convergence theorem using Theorem~\ref{thm-bounds.5}
for domination, we have, for any fixed compact set $X \subset \bR^{d}$
in momentum space and any $k$
\[
\int_{X} dp\ \vert \hat{\G}^{a}_{c,  * } (p) 
- \hat{\G}^{a}_{n} (p)\vert^{2} (1+p^{2})^{k}
\rightarrow 0
\]
as $n\rightarrow \infty$. This proves (\ref{eq-convergence.2}) of the
theorem, because we can choose $X$ to be a fixed Brillouin zone in the
dual of $(\e_{l}\bZ)^{d}$. The convergence in the
$L_{\infty}((\e_{l}\bZ)^{d})$ norm follows by Sobolev embedding (see
Appendix B).

\end{proof}

\section{Acknowledgements}

We thank Gianni Jona-Lasinio for his hospitality in Rome and for many
stimulating conversations over the years. PKM thanks G\'erard
Menessier and Andr\'e Neveu for fruitful discussions.  We
thank an anonymous referee for fingering an incorrect
Lemma.

\appendix
\section{Lattice Elliptic Regularity} \label{app-A}
\setcounter{equation}{0}

Suppose that $h$ solves 
\begin{equation}
(a-\Delta_{\e})h=g \label{eq-A.1}
\end{equation}  on $(\e \bZ)^{d}$
and $\varphi$ has compact support.  Then, on $(\e \bZ)^{d}$,
\[
\varphi h ( a -\Delta_{\e})h=\varphi h g
\]
Integrate over $(\e \bZ)^{d}$.  By the definition of $-\Delta_{\e}$ this
can be rewritten as
\begin{equation}
a\int_{(\e \bZ)^d}dz\ \varphi h^{2} +
\int_{(\e \bZ)^d}dz\ 
\sum_{\underline e \in  \Scal} (\nabla_{\underline e}\varphi h )
(\nabla_{\underline e}h ) = \int_{(\e \bZ)^d}dz\ \varphi h g
\label{eq-A.1.5}
\end{equation}
Surprisingly, calculation shows that this can be rewritten as
\begin{equation}\label{eq-A.1.51}
a\int_{(\e \bZ)^d}dz\ \varphi h^{2} + \frac{1}{2}
\sum_{\pm \underline e  \in \Scal }\int_{(\e \bZ)^d}dz\ 
\varphi (\nabla_{\underline e}h )^{2}
= \int_{(\e \bZ)^d}dz\ \big[\varphi h g
+ \frac{1}{2} (\Delta_{\e}\varphi)h^{2}\big]
\end{equation}
In other words  with forward and backward derivatives the lattice
gives the same formula as the continuum 
{\it without corrections that go to zero with $\e$}. 
For $a\ge 0$ and for $\varphi \ge 0$, (\ref{eq-A.1.51}) implies
\begin{equation}
\frac{1}{2} \sum_{\pm \underline e  \in \Scal }
\int_{(\e \bZ)^d}dz\ 
\varphi (\nabla_{\underline e}h )^{2}
\le \int_{(\e \bZ)^d}dz\ \big[\varphi h g
+ \frac{1}{2} (\Delta_{\e}\varphi)h^{2}\big]
\label{eq-A.2}
\end{equation}
\begin{equation}\label{eq-A.2a}
\int_{(\e \bZ)^d}dz\ \varphi h^{2}
\le \frac{1}{a}\int_{(\e \bZ)^d}dz\ \big[\varphi h g
+ \frac{1}{2} (\Delta_{\e}\varphi)h^{2}\big]
\end{equation}

Let
\[
[h]_{\varphi,j}^{2} = 2^{-j}
\sum_{ \pm \underline{e}_1,...,\pm
\underline{e}_j \in \Scal  }\int_{(\e \bZ)^{d}}dz\ \vert \nabla^{j}_{\underline
{e}_1,...,\underline {e}_j}h\vert^{2} \varphi
\]
By applying $j$ finite difference derivatives to (\ref{eq-A.1}) we find
that (\ref{eq-A.2}) is also true for $h$ and $g$ replaced by
derivatives of $h$ and $g$.  By the Cauchy-Schwartz inequality on the first
term in (\ref{eq-A.2}),
\[
[h]_{\varphi,j+1}^{2} \le [h]_{\varphi,j}[g]_{\varphi,j} +
[h]_{\frac{1}{2}\Delta_{\e}\varphi,j}^{2}
\]
Simplify the first term using the inequality $ab \le \frac{1}{2}
(a^{2}+b^{2})$ and use the resulting inequality to iteratively reduce
the order of the top derivative in the Sobolev norm,
\begin{equation}
\Vert h\Vert^{2}_{\varphi, k}:=\sum_{j=0}^{k}2^{-j} 
\sum_{\pm\underline
{e}_1,...,\pm\underline {e}_j \in \Scal }\int_{(\e \bZ)^{d}}dz\ \vert
\nabla^{j}_{\underline e_1,...,\underline e_j}h\vert^{2} \varphi
\label{eq-A.3}
\end{equation}
We obtain,
\begin{equation}
\Vert h\Vert^{2}_{\varphi_{k}, k} \le 
\Vert h\Vert^{2}_{\varphi_{0}, 0} 
+ \Vert g\Vert^{2}_{\varphi_{0}, k-1}
\label{eq-A.3.5}
\end{equation}
where $\varphi_{0}$ is the final member of a sequence $\varphi_{j}$ of
non-negative functions chosen such that
\begin{equation}\label{eq-A.3.55}
\varphi_{j-1} \ge \frac{1}{2}\varphi_{j} + 
\frac{1}{2}\Delta_{\e}\varphi_{j},
\ \ \varphi_{j-1}\ge \varphi_{j}
\end{equation}

\begin{proof} (Proposition~\ref{prop-bounds.4}). We are given
that $h$ solves (\ref{eq-A.1}) in $U_{\e} (R)$ with $g=0$. We can
estimate the $L_{2}$ norm in $U_{\e} (R/2)$ in two different
ways. Firstly, For all $a\ge 0$ we can use the maximum
principle $|h| \le \Vert f \Vert_{L^{\infty}(\partial
U_{\e}(R))}$
\[
\Vert h\Vert_{ L_{2} (U_{\e} (R/2)) } \le
cR^{d/2} \Vert f \Vert_{L^{\infty}(\partial U_{\e}(R))}
\]
Secondly, for $a\ge 1$, we can choose $\f$ in (\ref{eq-A.2a})
to be one on $U (R/2)$ and zero outside $U (R)$ to obtain
\begin{equation}\label{eq-A.37}
\Vert h\Vert^{2}_{L_{2} (U_{\e} (R/2))}
\le
C_{R} a^{-1}
\Vert h\Vert^{2}_{L_{2} (U_{\e} (R))}
\end{equation}
and then use the maximum principle to bound the right hand side by the
$L^{\infty}(\partial U_{\e}(R))$ norm.  Therefore, for any smooth
$\f_{0}$ supported in $U_{\e} (R/2)$
\begin{equation}\label{eq-A.37a}
\Vert h\Vert^{2}_{\f_{0},0} \le
C_{R} (1+a)^{-1}\Vert f \Vert_{L^{\infty}(\partial U_{\e}(R))}^{ 2}
\end{equation}
Let $\varphi_{k}, \varphi_{k-1},\dotsc ,\f_{0} \ge 0$ be $C^{\infty}$
continuum functions with compact support in $U ( R/2)$ such that
$\varphi_{k}=1$ on $\Omega$ and (\ref{eq-A.3.55}) holds.  Apply
(\ref{eq-A.3.5}) and (\ref{eq-A.37a}) to obtain
Proposition~\ref{prop-bounds.4}.
\end{proof}

\emph{Remark on Exponential Decay:} The correct $a$
dependence is $\exp (-O (\sqrt{a}R))$.  We outline how to do this
using a method suggested by \cite{Agmon82}. (\ref{eq-A.2a}), for
$g=0$, can be rewritten as
\begin{gather*}
\int_{(\e \bZ)^d}dz\ w h^{2} \le 0\\
w:= (a-\frac{1}{2}\Delta_{\e})\varphi
\end{gather*}
Consider the choice $\f=\exp (-u)$ with
\[
u(x) = \frac{1}{2}\sqrt{a}\sqrt{1+x^{2}}
\]
Let $\D_{0}$ be the continuum Laplacian. The finite difference
Laplacian $\D_{\e}\f$ can be written as an integral over $\D_{0}\f$: for
example, in one dimension
\[
\D_{\e}\f (x) = \int_{0}^{1}\int_{0}^{1} ds\ dt\ \D_{0}\f(x+[t-s]\e)
\]
Using this we find $\D_{\e} u \ge 0$ and $w \ge c (a,\e)\f$ with $c
(a,\e)\approx \exp (-\sqrt{a}\e)$.  Now replace $\exp (-u)$ by
$\f=\exp (-u)\psi$ where  $\psi$ is a smooth, positive, monotonic
decreasing function such that \ 
$\psi=1$ on  $U(\frac{2}{3}R)$  and vanishes outside 
$U(R)$.  Then $w\ge c (a,\e)\f \ge 0$ on  $U(\frac{2}{3}R)$ 
so by taking the part
of the integral where $w \not \ge 0$ to the right hand side of the
bound and discarding part of the integral where $w \ge 0$ we get 
\[
c (a,\e)\int_{ U_{\e} (R/2)}dz\ \f h^{2} 
\le \int_{ U_{\e} (R)\setminus U_{\e}
(\frac{2}{3}R) }dz\ (-w) h^{2} \le \Vert f\Vert^{2}_{L_{\infty} 
(\partial U_{\e}
(R))}\int_{ U_{\e}(R) \setminus U_{\e}( \frac{2}{3}R)}
 dz\ \vert w\vert
\]
 which gives decay because 
\[
\f \vert_{{U (R/2)}} \ge   \exp ( O
(\sqrt{a}R)) \f\vert_{U (R)\setminus U (\frac{2}{3}R)}
\]

\emph{Preparation for proof of
Lemma~\ref{lem-convergence.4}:} \  Suppose (\ref{eq-A.1})
holds in a domain $U_{\e} (R)$ and $h$ vanishes on $\partial U_{\e}
(R)$. Then, with $\varphi =1$,  $a \ge 0$, 
(\ref{eq-A.1.5}) becomes,
\begin{equation}
\sum_{\underline e} \int_{U_{\e} (R)}dz\ 
(\nabla_{\underline e} h )
(\nabla_{\underline e}h )
\le  \int_{U_{\e} (R)}dz\ h g
\label{eq-A.4}
\end{equation}
We estimate the right hand side by the Cauchy-Schwartz inequality
and substitute the result into the Poincar\'e inequality, which is
\begin{equation}
\Vert h\Vert^{2}_{L^{2} (U_{\epsilon} (R))}
\le C_{R} 
\sum_{\underline e} \int_{U_{\e}}dz\ 
(\nabla_{\underline e} h )
(\nabla_{\underline e}h )
\label{eq-A.5}
\end{equation}
where $C_{R}$ is independant of $\e$. Then
\begin{equation}
\Vert h\Vert_{L^{2} (U_{\epsilon} (R))}
\le
C_{R} \Vert g\Vert_{L^{2} (U_{\epsilon} (R))}
\label{eq-A.6}
\end{equation}
Returning to (\ref{eq-A.3.5}) and using (\ref{eq-A.6}) we obtain
\begin{lemma}\label{lem-A.1}
Let $\varphi \ge 0$ be a $C^{\infty}$ continuum function with compact
support in $U (R)$.  By restriction it defines functions on all
lattices, denoted by the same letter. Then for $\epsilon$ sufficiently
small, there exists a constant $C_{R,\varphi }$ such that the solution to
(\ref{eq-A.1}), with zero boundary conditions on $\partial U_{\e} (R)$,
satisfies 
\[
\Vert h\Vert_{\varphi, k} \le C_{R,\phi} \Vert g\Vert_{H_{k-1}
(U_{\epsilon} (R))}
\]
where the constant $ C_{ R,\phi}$ is independant of
$\epsilon$.
\end{lemma}

\begin{proof} (Lemma~\ref{lem-convergence.4}) Write as in {\it
lemma 4.3}, $ -\D_{\e_{n+1}} = -\D_{\e_{n}} +Q$.  Let
$h=h^{a}_{\e_{n}}- h^{a}_{\e_{n+1}}$.  This has zero boundary
conditions on ${\partial U}_{\e_{n}}(R)$ and
\[
(-\D_{\e_{n}} + a)h(x)=g(x)\ :\ x\in U_{\e_{n}}(R)
\]
where $g=Q h^{a}_{\e_{n+1}}$.  Now apply Lemma~\ref{lem-A.1} and
estimate $g$ by Lemma~\ref{lem-convergence.3} followed by
Proposition~\ref{prop-bounds.4}. The proof is completed by Sobolev
embedding, see Lemma~\ref{lem-B.1} below, taking $k$ in
Lemma~\ref{lem-A.1} sufficiently large.
\end{proof}

\section{Sobolev Spaces on the Lattice}\label{app-B}
\setcounter{equation}{0}
 
\begin{lemma}\label{lem-B.1}
Let $I^{d} := [0,1]^{d}$, $I^{d}_{\e} = I^{d}\cap (\e
\bZ)^d$ and $X$ be any subset of $I^{d}_{\e}$.  Then
\[
 \Vert u\Vert_{L^{\infty}(X)} \le C_{d}
\Vert u\Vert_{H_{d}(I^{d}_{\e})}
\]
\end{lemma}

\begin{proof} \ \ A smooth continuum function $u (x)$,
where $x = (x_{1},\dots ,x_{d})$, satisfies
\[
x_{1}\dots x_{d}u (x) =
\int_{0}^{x_{1}}dy_{1}\dots\int_{0}^{x_{d}}dy_{d} \,\partial_{1}\dots
\partial_{d} (y_{1}\dots y_{d}u(y_{1},\dots ,y_{d}))
\]
and therefore, for $x \in I^{d} := [0,1]^{d}$, 
\[
|u (x)| \le \frac{C_d}{|x_{1}\dots x_{d}|}\Vert u\Vert_{H_{d} (I^{d})}
\]
The same proof adapts to the lattice $(\e \bZ)^d$ with integrals and
derivatives being replaced by sums and finite differences so that for
$u:I^{d}_{\e} \rightarrow \bR$, 
\[
\sup_{x \in J_{\e}} |u (x)| \le C_{d} \Vert u\Vert_{H_{d}
(I^{d}_{\e})}
\]
where $J_{\e}:=[1/2,1]^{d}_{\e}$.  For $d=2$, $I^{2}$ is the union of
$[1/2,1]^{2}$, $[0,1/2]^{2}$, $[1/2,1]\times [0,1/2]$ and
$[0,1/2]\times [1/2,1]$ and by symmetry the same bound holds with
$J_{\e}$ replaced by any of these boxes. The same argument applies for
all dimensions $d$, which implies the lemma. \end{proof}

\begin{lemma}[Poincare Inequality]\label{lem-B.2}
Let $u:I^{d}_{\e}\rightarrow \bR$ be any function
vanishing on the boundary of $I^{d}_{\e}$. There exists $C$
independent of $\e$ such that
\[
\int_{I^{d}_{\e}} \, dx |u|^{2} \le C\Vert \nabla u\Vert^{2}_{L^{2}
(I^{d}_{\e})}
\]
\end{lemma}

\begin{proof}\ \ Let $u_{i} (x)$ be the finite difference partial
derivative with respect to component $x_{i}$ of $x$.  Then, for $i =
1,\dots ,d$,
\[
|u (x_{1},\dots ,x_{d})| \le \int_{I_{\e}} \, dx_{i} |u_{i}
(x_{1},\dots ,x_{d}) |
\]
The right hand side is a function of all components of $x$ except
$x_{i}$. Take the product over $i$ followed by $2/d$ root or power.
\[
|u (x)|^{2} \le \prod_{i} \bigg(\int_{I_{\e}} \, dx_{i}
|u_{i}|\bigg)^{\frac{2}{ d}} \le \prod_{i} \bigg(\int_{I_{\e}} \,
dx_{i} |u_{i}|^{2}\bigg)^{\frac{1}{ d}}
\]
Integrate both sides over $(x_{1},\dots ,x_{d})\in I^{d}_{\e}$ and use
the H\"older inequality on the right hand side.
\[
\int_{I^{d}_{\e}} \, dx |u|^{2} \le \prod_{i} \bigg(\int_{I^{d}_{\e}}
\, dx \int_{I_{\e}} \, dx_{i} |u_{i}|^{2}\bigg)^{\frac{1}{ d}} \le
\int_{I^{d}_{\e}} \, dx \sum_{i}|u_{i}|^{2}
\]
because the extra integral integrates to unity. \end{proof}


\begin{thebibliography}{Agm82}

\bibitem[Agm65]{Agmon65}
Shmuel Agmon.
\newblock {\em Lectures on elliptic boundary value problems}.
\newblock Prepared for publication by B. Frank Jones, Jr. with the assistance
  of George W. Batten, Jr. Van Nostrand Mathematical Studies, No. 2. D. Van
  Nostrand Co., Inc., Princeton, N.J.-Toronto-London, 1965.

\bibitem[Agm82]{Agmon82}
Shmuel Agmon.
\newblock {\em Lectures on exponential decay of solutions of second-order
  elliptic equations: bounds on eigenfunctions of {$N$}-body {S}chr\"odinger
  operators}, volume~29 of {\em Mathematical Notes}.
\newblock Princeton University Press, Princeton, NJ, 1982.
 
\bibitem[Ba{\l}82a]{Bal82a}
Tadeusz Ba{\l}aban.
\newblock ({H}iggs){$\sb{2,3}$} quantum fields in a finite volume. {I}. {A}
  lower bound.
\newblock {\em Commun. Math. Phys.}, 85(4):603--626, 1982.

\bibitem[Ba{\l}82b]{Bal82b}
Tadeusz Ba{\l}aban.
\newblock $({H}iggs)_{2,3}$ quantum fields in a finite volume {II}. {A}n upper
  bound.
\newblock {\em Commun. Math. Phys.}, 86:555--594, 1982.

\bibitem[BMS]{BMS03}
D.C Brydges, P.K. Mitter, and B.~Scoppola.
\newblock Critical $({\bf\Phi}^{4})_{3,\>\epsilon}$. 
\newblock {\em hep-th}/0206040, {\em mp-arc} 02-273, {\em Commun. Math. Phys.},
(in press).

\bibitem[FS81]{FrSp81}
J.~Fr\"ohlich and T.~Spencer.
\newblock {K}osterlitz-{T}houless transition in the two dimensional {A}belian
  spin systems and {C}oulomb gas.
\newblock {\em  Comm. Math. Phys. }, 81:527, 1981.

\bibitem[GK80]{GaKu80}
K.~Gawedzki and A.~Kupiainen.
\newblock A rigorous block spin approach to massless lattice theories.
\newblock {\em  Comm. Math. Phys. }, 77:31--64, 1980.

\bibitem[GK83]{GaKu83}
K.~Gawedzki and A.~Kupiainen.
\newblock Block spin renormalization group for dipole gas and
  $(\bigtriangledown \phi)^4$.
\newblock {\em Ann. Phys.}, 147:198, 1983.

\bibitem[GK86]{GaKu86}
K.~Gawedzki and A.~Kupiainen.
\newblock Asymptotic freedom beyond perturbation theory.
\newblock In K.~Osterwalder and R.~Stora, editors, {\em Critical Phenomena,
  Random Systems, Gauge Theories}. Les Houches, North Holland, 1986.

\bibitem[HS02]{HaiSei2002}  
Christian Hainzl and Robert Seiringer.
\newblock General decomposition of radial functions on {$\mathbb R\sp n$} and
  applications to {$N$}-body quantum systems.
\newblock {\em Lett. Math. Phys.}, 61(1):75--84, 2002. 

\bibitem[MS00]{MiSc2000}
P.~K. Mitter and B.~Scoppola.
\newblock Renormalization group approach to interacting polymerised manifolds.
\newblock {\em Comm. Math. Phys.}, 209(1):207--261, 2000.

\end{thebibliography}

\end{document}